\documentclass[aps,prd,preprintnumbers,notitlepage,superscriptaddress,nofootinbib]{revtex4-1}%showpacs
\usepackage{ulem}
\usepackage[dvips]{graphicx}
\usepackage{bm,latexsym,amsmath,amssymb,amsfonts}
\usepackage[colorlinks=true,linkcolor=magenta,citecolor=magenta]{hyperref}
\newcommand*{\ba}{\begin{eqnarray}}
	\newcommand*{\ea}{\end{eqnarray}}

\newcommand{\simgt}{\lower.5ex\hbox{$\; \buildrel > \over \sim \;$}}
\newcommand{\simlt}{\lower.5ex\hbox{$\; \buildrel < \over \sim \;$}}
\newcommand*{\eps}{\varepsilon}
\newcommand*{\Lag}{{\cal L}}

\newcommand*{\p}{\partial}

\newcommand*{\psib}{{\bar \psi}}
\newcommand*{\alphad}{{\dot \alpha}}
\newcommand*{\betad}{{\dot \beta}}
\newcommand*{\gammad}{{\dot \gamma}}
\newcommand*{\deltad}{{\dot \delta}}

\newcommand*{\Psib}{{\bar \Psi}}

\newcommand*{\PI}{P^{(1)}}
\newcommand*{\PII}{P^{(2)}}
\newcommand*{\PIII}{P^{(3)}}

\newcommand*{\SI}{S^{(1)}}
\newcommand*{\SII}{S^{(2)}}
\newcommand*{\SIII}{S^{(3)}}
\newcommand*{\SIIII}{S^{(4)}}

\newcommand*{\bx}{{\bf x}}
\newcommand*{\by}{{\bf y}}
\newcommand*{\bz}{{\bf z}}
\newcommand*{\phid}{{\dot \phi}}
\newcommand*{\phidd}{{\ddot \phi}}
\newcommand*{\psid}{{\dot \psi}}
\newcommand*{\psibd}{{\dot \psib}}
\newcommand*{\psidd}{{\ddot \psi}}
\newcommand*{\psibdd}{{\ddot \psib}}
\newcommand*{\thetad}{{\dot \theta}}

\newcommand*{\B}{{\cal B}}
\newcommand*{\C}{{\cal C}}
\newcommand*{\E}{{\cal E}}
\newcommand*{\X}{{\cal X}}
\newcommand*{\Y}{{\cal Y}}
\newcommand*{\F}{{\cal F}}
\newcommand*{\G}{{\cal G}}

\newcommand*{\sigmab}{{\bar \sigma}}

\def\({\Bigl(}
\def\){\Bigr)}
\def\[{\Biggl[}
\def\]{\Biggr]}

\begin{document}

\title{Ghost-free scalar-fermion interactions}

\author{Rampei~Kimura}
\email[Email: ]{rampei"at"aoni.waseda.jp}
\affiliation{Waseda Institute for Advanced Study, Waseda University, 1-6-1 Nishi-Waseda, Shinjuku, Tokyo 169-8050, Japan}
\affiliation{Department of Physics, Tokyo Institute of Technology, Tokyo
152-8551, Japan}

\author{Yuki~Sakakihara}
\email[Email: ]{y.sakakihara"at"kwansei.ac.jp}
\affiliation{Department of Physics, Kwansei Gakuin University, Sanda, Hyogo 669-1337, Japan}

\author{Masahide~Yamaguchi}
\email[Email: ]{gucci"at"phys.titech.ac.jp}
\affiliation{Department of Physics, Tokyo Institute of Technology, Tokyo
152-8551, Japan}

\begin{abstract}
We discuss a covariant extension of interactions between scalar fields
and fermions in a flat space-time.  We show, in a covariant theory, how
to evade fermionic ghosts appearing because of the extra degrees of
freedom behind a fermionic nature even in the Lagrangian with first
derivatives.  We will give a concrete example of a quadratic theory with
up to the first derivative of multiple scalar fields and a Weyl fermion.  We
examine not only the maximally degenerate condition, which makes the
number of degrees of freedom correct, but also a supplementary condition
guaranteeing that the time evolution takes place properly.  We also show
that proposed derivative interaction terms between scalar fields and a
Weyl fermion cannot be removed by field redefinitions.
\end{abstract}

%\pacs{}
\preprint{}
\maketitle
%%%%%%%%%%%%%%%%%%%%%%%%%%%%%%%%%%%%%%%%%%%%%%%%%%%%%
\section{Introduction}
%%%%%%%%%%%%%%%%%%%%%%%%%%%%%%%%%%%%%%%%%%%%%%%%%%%%%

Construction of general theory without ghost degrees of freedom (d.o.f.) has been
discussed for a long time.  According to Ostrogradsky's theorem, a
ghost always appears in a higher (time) derivative theory as long as it
is nondegenerate \cite{Ostrogradsky:1850fid} (see also
Ref.~\cite{Woodard:2015zca}). If a term with higher derivatives plays an
important role in the dynamics, the ghost d.o.f. associated
with it must be removed because, otherwise, the dynamics is
unstable. This is a different approach from effective field
theory, which allows a ghost as long as it appears above the scale
we are interested in; that is, a term with higher derivatives can be
treated as a perturbation.

Such an Ostrogradsky ghost could be circumvented by introducing the
degeneracy of the kinetic matrix, which leads to the existence of a
primary constraint and a series of subsequent constraints, which are
responsible for eliminating extra ghost d.o.f. associated
with higher derivatives. After the recent rediscovery
\cite{Charmousis:2011bf,Deffayet:2011gz,Kobayashi:2011nu} of Horndeski
theory \cite{Horndeski:1974wa} in the context of the Galileon
\cite{Nicolis:2008in}, the pursuit of finding general theory without such
ghost d.o.f. has been revived, especially for bosonic
d.o.f., in the context of point
particles~\cite{Motohashi:2016ftl,Klein:2016aiq,Motohashi:2017eya,Motohashi:2018pxg},
their field theoretical application~\cite{Crisostomi:2017aim},
scalar-tensor theories~\cite{Horndeski:1974wa, Deffayet:2011gz,
Kobayashi:2011nu, Padilla:2012dx, Zumalacarregui:2013pma,
Gleyzes:2014dya, Gao:2014soa, Langlois:2015cwa, Ohashi:2015fma,
Crisostomi:2016czh, BenAchour:2016fzp}, vector-tensor
theories~\cite{Horndeski:1976gi, Deffayet:2013tca, Heisenberg:2014rta,
Tasinato:2014mia,Tasinato:2014eka,
Allys:2015sht,Jimenez:2016isa,Heisenberg:2016eld, Kimura:2016rzw}, and
form fields~\cite{Deffayet:2010zh, Deffayet:2016von, Deffayet:2017eqq}.

 On the other hand, generic theory with fermionic d.o.f.
has not yet been investigated well. If we could find such fermionic
theories which have not been explored, some applications of them will be
expected. One will be fermionic dark matter, the interaction of which could be
constrained by the condition of the absence of ghost d.o.f. Another application is inflation and its subsequent reheating,
which needs interactions between an inflaton and standard model
particles, since their interactions are still unknown. The new
interaction may affect the history during the reheating era through particle
production. The effect of the interactions can also appear in
observables such as non-Gaussianinty through the loop corrections of standard model
particles, as pointed out in Refs.~\cite{Chen:2016nrs, Chen:2016uwp}. The
construction of a ghost-free (higher derivative) supersymmetric theory
is another direction, though some attempts have already been made
\cite{Khoury:2011da,Farakos:2013zya,Kimura:2016irk,Fujimori:2017kyi}.  

What is prominent in fermionic theory is that ghosts appear even
with no higher derivatives in the usual meaning, i.e., no second
derivatives and higher.  As mentioned in Ref.~\cite{Henneaux:1992ig} for
a purely fermionic system, if $N$ fermionic variables carry $2N$ d.o.f. in the phase space, then negative norm states inevitably
appear. That is why we usually see that the canonical kinetic term of
the fermionic field linearly depends on the derivative of the field like
the Dirac Lagrangian. Indeed, in the case of usual Weyl fermions, such unwanted
states are evaded by the existence of a sufficient number of primary
constraints.  In our previous paper \cite{Kimura:2017gcy}, as a starting
point, we studied point particle theories, the Lagrangian of which contains
both bosonic and fermionic variables with their first derivatives,
and showed that the coexistence of fermionic d.o.f. and
bosonic ones allows us to have some extension of their kinetic terms.
In the present paper, we apply the analysis done in
Ref.~\cite{Kimura:2017gcy} to the field system with scalar fields and Weyl
fermions, which can be even a preliminary step toward tensor-fermion
or any other interesting theories including fermionic d.o.f.

This paper is organized as follows. In Sec.~\ref{sec:II}, following
the previous analysis for point particle theories, we give a
general formulation of the construction of covariant theory with up to
first derivatives of $n$-scalar fields and $N$-Weyl fermions.  We then derive $4N$
primary constraints by introducing maximally degenerate conditions.
In Sec.~\ref{sec:III}, we concretely write down the quadratic theory of
$n$-scalar and one Weyl fermion fields and apply the set of the
maximally degenerate conditions, which we have proposed for removing the
fermionic ghosts properly. In Sec.~\ref{sec:IV}, we perform Hamiltonian
analysis of the quadratic theory and derive a supplementary condition
such that all the primary constraints are second class. In
Sec.~\ref{sec:V}, we obtain the explicit counterparts in Lagrangian
formulation to the conditions in Hamiltonian formulation.  In
Sec.~\ref{sec:VI}, we show that the obtained theories satisfying the
maximally degenerate conditions cannot be mapped into theories of which the 
Lagrangian linearly depends on the derivative of the fermionic fields.
In Sec.~\ref{sec:summary}, we give a summary of our work.  In
Appendix~\ref{app:A}, we summarize definitions and identities of the
Pauli matrices.  In Appendix~\ref{app:B}, we show the equivalence
between the maximally degenerate conditions and the primary constraints
obtained in Sec.~\ref{sec:II}.  In Appendix~\ref{app:C}, we extend our
analysis in Sec.~\ref{sec:III} to multiple Weyl fermions and derive the
primary constraints.

%%%%%%%%%%%%%%%%%%%%%%%%%%%%%%%%%%%%%
\section{General formalism for constructing degenerate Lagrangian}
\label{sec:II}
%%%%%%%%%%%%%%%%%%%%%%%%%%%%%%%%%%%%%

In the present paper, we construct flat space-time theories with $n$ real
scalar fields $\phi^a (t, \bx)$ $(a=1,..., n)$, $N$ Weyl fermionic
fields $\psi_I^\alpha (t, \bx)$ and their Hermitian
conjugates~$\psib_I^\alphad (t, \bx)$ $(\alpha, \alphad = 1 \textrm{ or
} 2,$ $I=1,..., N)$.  Following our previous work
\cite{Kimura:2017gcy}, we consider the Lagrangian of which the fields carry up
to the first derivative.  The most general action is symbolized by
\ba
S = \int d^4x \, \Lag\left[\phi^a, \p_\mu \phi^a, \psi_I^\alpha, \p_\mu \psi_I^\alpha, \psib_I^\alphad, \p_\mu \psib_I^\alphad\right] \,.
\label{Lagrangian}
\ea
Here, the Lorentzian indices are raised and lowered by the Minkowski
metric $\eta_{\mu\nu}$, and the fermionic indices are raised and lowered
by the antisymmetric tensors $\eps_{\alpha\beta}$ and
$\eps_{\alphad\betad}$.  In addition, the capital latin indices $(I, J,
K, ...)$ are contracted with the Kronecker delta $\delta_{IJ}$ such as
$\psi^\alpha_I \psi_{I,\alpha}:=\delta_{IJ}\psi^\alpha_I
\psi_{J,\alpha}$.  Throughout this paper, we use the metric signature
$(+,-,-,-)$.

When the Lagrangian (\ref{Lagrangian}) consists only of scalar
fields, the absence of Ostrogradsky's ghosts is automatically
ensured since the Euler-Lagrange equations contain no more than
second derivatives with respect to time.  On the other hand, second
derivatives in fermionic equations of motion are generally dangerous
because extra d.o.f. in the fermionic sector immediately lead to negative
norm states (see Refs.~\cite{Henneaux:1992ig,Kimura:2017gcy} for the detail.).
In order to avoid such ghost d.o.f., the system must contain the
appropriate number of constraints, which originate from the degeneracy
of the kinetic matrix. In this section, we derive degeneracy conditions
and a supplementary condition for the Lagrangian (\ref{Lagrangian}).
The basic
treatment of Grassmann algebra and Hamiltonian formulation, of which we
are making use, are summarized in, e.g.,
Refs.~\cite{Henneaux:1992ig,Kimura:2017gcy}, and we use the left derivative
throughout this paper.

We find degeneracy conditions of \eqref{Lagrangian}, which yield an appropriate number of primary constraints, eliminating half of the d.o.f. of fermions in phase space. For deriving the conditions, we first take a look at the canonical momenta defined as 
\ba
\pi_{\phi^a}^\mu = {\p \Lag \over \p (\p_\mu \phi^a)}, \qquad
\pi_{\psi_I^\alpha}^\mu = {\p \Lag \over \p (\p_\mu \psi_I^\alpha)}, \qquad
\pi_{\psib_I^\alphad}^\mu = {\p \Lag \over \p (\p_\mu \psib_I^\alphad)}
= -\bigl( \pi_{\psi_I^\alpha}^\mu \bigr)^{\dagger} \,. 
\ea
Then, the variations of the $0$th component of canonical momenta with respect to all canonical variables 
give the following set of equations, 
\ba
\begin{pmatrix}
	\delta \pi_{\phi^a} \\
	\delta \pi_{\psi_I^\alpha} \\
	\delta \pi_{\psib_I^\alphad}
\end{pmatrix}
=
{\bf K}
\left(
\begin{array}{c}
	\delta \phid^b  \\
	\delta \psid_J^\beta  \\
	\delta \psibd{}_J^\betad  \\
\end{array}
\right)+
\left(
\begin{array}{c}
	\delta z_a^{(\phi)}  \\
	\delta z_{\alpha, I}^{(\psi)} \\
	\delta z_{\alphad,I}^{(\psib)} \\
\end{array}
\right)\,,
\label{infmom}
\ea
where we have omitted the superscript of the $0$th component, i.e., $\pi_A\equiv\pi^0_A$, and defined the kinetic matrix,
\ba
{\bf K}&=&
\left(
\begin{array}{ccc}
	A_{ab} & \B_{a\beta,J} & \B_{a\betad,J}\\
	\C_{\alpha b,I} & D_{\alpha\beta,IJ} & D_{\alpha\betad,IJ}\\
	\C_{\alphad b,I} & D_{\alphad\beta,IJ} & D_{\alphad\betad,IJ}\\
\end{array}
\right)
=\left(
\begin{array}{ccc}
	\Lag_{\phid^a \phid^b} & -\Lag_{\phid^a \psid_J^\beta} & -\Lag_{\phid^a \psibd{}_J^\betad}\\
	\Lag_{\psid_I^\alpha \phid^b} & ~~\Lag_{\psid_I^\alpha  \psid_J^\beta} & \Lag_{\psid_I^\alpha  \psibd{}_J^\betad}\\
	\Lag_{\psibd{}_I^\alphad  \phid^b} & ~~\Lag_{\psibd{}_I^\alphad \psid_J^\beta} & \Lag_{\psibd{}_I^\alphad \psibd{}_J^\betad}\\
\end{array}
\right), 
\label{KineticMatrix}
\ea
and
\ba
\left(
\begin{array}{c}
	\delta z_a^{(\phi)}  \\
	\delta z_{\alpha,I}^{(\psi)} \\
	\delta z_{\alphad,I}^{(\psib)} \\
\end{array}
\right)
&=&
\left(
\begin{array}{ccc}
	\Lag_{\phid^a \p_i \phi^b} & -\Lag_{\phid^a \p_i\psi_J^\beta} & -\Lag_{\phid^a \p_i\psib{}_J^\betad}\\
	\Lag_{\psid_I^\alpha \p_i\phi^b} & ~~\Lag_{\psid_I^\alpha  \p_i\psi_J^\beta} & \Lag_{\psid_I^\alpha  \p_i\psib{}_J^\betad}\\
	\Lag_{\psibd{}_I^\alphad  \p_i\phi^b} & ~~\Lag_{\psibd{}_I^\alphad \p_i\psi^\beta} & \Lag_{\psibd{}_I^\alphad \p_i\psib{}_J^\betad}\\
\end{array}
\right)
\left(
\begin{array}{c}
	\delta (\p_i \phi^b)  \\
	\delta (\p_i \psi_J^\beta)  \\
	\delta (\p_i \psib_J^\betad)  \\
\end{array}
\right)
+
\left(
\begin{array}{ccc}
	\Lag_{\phid^a  \phi^b} & -\Lag_{\phid^a \psi_J^\beta} & -\Lag_{\phid^a \psib{}_J^\betad}\\
	\Lag_{\psid_I^\alpha \phi^b} & ~~\Lag_{\psid_I^\alpha  \psi_J^\beta} & ~~\Lag_{\psid_I^\alpha  \psib{}_J^\betad}\\
	\Lag_{\psibd{}_I^\alphad \phi^b} & ~~\Lag_{\psibd{}_I^\alphad\psi_J^\beta} & ~~\Lag_{\psibd{}_I^\alphad \psib{}_J^\betad}\\
\end{array}
\right)
\left(
\begin{array}{c}
	\delta \phi^b  \\
	\delta \psi_J^\beta  \\
	\delta \psib_J^\betad \\
\end{array}
\right)\,.
\ea
Here, we have introduced the shortcut notation, 
\begin{align}
	\Lag_{XY}=\frac{\partial^2 \Lag}{\partial Y\partial X}=\frac{\partial }{\partial Y}\Bigl(\frac{\partial \Lag}{\partial X}\Bigr)\ .
\end{align} 
Here, $A_{ab} $ is a symmetric matrix, while $D_{\alpha\beta,IJ}$, $D_{\alpha\betad,IJ}$, $D_{\alphad\beta,IJ}$, and $D_{\alphad\betad,IJ}$ are antisymmetric matrices under the exchange of the greek indices as 
\begin{align}
  D_{\alpha\beta,IJ}=-D_{\beta\alpha,JI}\ , \quad  D_{\alpha\betad,IJ}=-D_{\betad\alpha,JI} \ , \quad  D_{\alphad\betad,IJ}=-D_{\betad\alphad,JI} \ ,
\end{align}
and ${\cal B}$ and ${\cal C}$ are Grassmann-odd matrices and related as ${\cal C}_{\alpha b, I}=-{\cal B}_{b \alpha, I}$ and ${\cal C}_{\alphad b, I}=-{\cal B}_{b \alphad, I}$.  
The Hermitian properties of $\pi_{\phi^a}^\mu$ and the anti-Hermitian properties of $\pi_{\psi^\alpha_I}^\mu$, i.e., $(\pi_{\phi^a}^\mu)^\dagger = \pi_{\phi^a}^\mu$ and $(\pi_{\psi_I^\alpha}^\mu)^\dagger = -\pi_{\psib_I^\alphad}^\mu$, lead to 
\ba
A_{ab}=A_{ab}^\dagger, \quad
\B_{a\beta,J}=-\B_{a\betad,J}^\dagger, \quad
\C_{\alpha b,I}=-\C_{\alphad b,I}^\dagger, \quad
D_{\alphad \beta,IJ}=-D_{\alpha \betad,IJ}^\dagger, \quad
D_{\alpha \beta,IJ}=-D_{\alphad \betad,IJ}^\dagger \,.
\ea

In the present paper, we assume that the scalar submatrix of the kinetic matrix $A_{ab}$ is nondegenerate, i.e., invertible. This assumption is equivalent to requiring
\begin{align}
	\det A_{ab}^{(0)} \neq 0 \ , 
\end{align}
where $\det A_{ab}^{(0)}$ is defined by setting all fermionic variables and their derivatives to zero. 
Then, the first equation (\ref{infmom}) can be solved for $\delta \phid^b$,
\ba
\delta\dot{\phi}{}^b=A^{ba} \left(
 \delta \pi_{\phi^a} - B_{a\beta,J} \delta \dot{\psi}{}_J^\beta - B_{a\betad,J}\delta \dot{\psib}{}_J^\betad - \delta z_a^{(\phi)}
\right)\,,
\ea
where we have defined the inverse of the kinetic matrix in the scalar sector as
$A^{ab} = (A^{-1})^{ab}$. 
Plugging this into the infinitesimal momenta of $\psi^\alpha$ and $\psib^\alphad$,  we get
\ba
\delta \pi_{\psi_I^\alpha} &=& 
\(D_{\alpha\beta,IJ}-\C_{\alpha b,I} A^{ba} \B_{a\beta,J}\)\delta\dot{ \psi}{}_J^\beta
+\(D_{\alpha\betad,IJ}-\C_{\alpha b,I} A^{ba} \B_{a\betad,J}\)\delta \dot{\psib}{}_J^\betad 
+C_{\alpha b,I} A^{ba} \(\delta \pi_{\phi^a} -\delta z_a^{(\phi)}\) +\delta z_{\alpha,I}^{(\psi)}\,,\\
\delta \pi_{\psib_I^\alphad} &=&
\(D_{\alphad\beta,IJ}-\C_{\alphad b,I} A^{ba} \B_{a\beta,J}\)\delta \dot{\psi}{}_J^\beta
+\(D_{\alphad\betad,IJ}-\C_{\alphad b,I} A^{ba} \B_{a\betad,J}\)\delta \dot{\psib}{}_J^\betad 
+C_{\alphad b,I} A^{ba} \(\delta \pi_{\phi^a} -\delta z_a^{(\phi)}\) +\delta z_{\alphad,I}^{(\psib)}\,.
\ea
In order to obtain the sufficient number of constraints, 
we need conditions
that 
the velocity terms $\delta\dot{\psi_I}$ and $\delta\dot{\psib}_I$
cannot be expressed in terms of the canonical variables. 
In the present paper, we adopt the ``maximally degenerate conditions,''\footnote{
There should be other candidates for conditions eliminating ghost d.o.f., 
but here we adopt the simplest case, in which all the constraints eliminating extra d.o.f.
are primary constraints. See Ref.~\cite{Kimura:2017gcy} for the other possibilities. 
} i.e.,
\ba
&&D_{\alpha\beta,IJ}-\C_{\alpha b,I} A^{ba} \B_{a\beta,J}=0, \qquad D_{\alpha\betad,IJ}-\C_{\alpha b,I} A^{ba} \B_{a\betad,J}=0\,, \label{MDC1}\\
&&D_{\alphad\beta,IJ}-\C_{\alphad b,I} A^{ba} \B_{a\beta,J}=0, \qquad D_{\alphad\betad,IJ}-\C_{\alphad b,I} A^{ba} \B_{a\betad,J}=0\,. \label{MDC2}
\ea
Two of these four conditions are equivalent to the others since
they are related through Hermitian conjugates, 
\ba
\(D_{\alpha\beta,IJ}-\C_{\alpha b,I} A^{ba} \B_{a\beta,J}\)^\dagger&=&-\(D_{\alphad\betad,IJ}-\C_{\alphad b,I} A^{ba} \B_{a\betad,J}\)\,,\\
\(D_{\alpha\betad,IJ}-\C_{\alpha b,I} A^{ba} \B_{a\betad,J}\)^\dagger&=&-\(D_{\alphad\beta,IJ}-\C_{\alphad b,I} A^{ba} \B_{a\beta,J}\)\,.
\ea
The maximally degenerate conditions lead to the following $4N$ primary constraints,
\ba
\Phi_{\psi_I^\alpha} &\equiv& 
\pi_{\psi_I^\alpha} - F_{\alpha, I} (\phi^a, \pi_{\phi^a}, \p_i \phi^a, \psi_J^\beta, \p_i \psi_J^\beta, \psib_J^\betad, \p_i \psib_J^\betad) =0 \,,\label{PRI1}\\
\Phi_{\psib_I^\alphad} &=& - \left( \Phi_{\psi_I^\alpha} \right)^{\dagger} =  \pi_{\psib_I^\alphad}- G_{\alphad,I} (\phi^a, \pi_{\phi^a}, \p_i \phi^a, \psi_J^\beta, \p_i \psi_J^\beta, \psib_J^\betad, \p_i \psib_J^\betad) =0 \,,\label{PRI2}
\ea
with $G_{\alphad,I} = - (F_{\alphad,I})^{\dagger}$. The explicit proof of the equivalence between Eqs.~(\ref{PRI1}) and (\ref{PRI2}) and Eqs.~(\ref{MDC1}) and (\ref{MDC2})
is shown in Appendix~\ref{app:B}.
As discussed in Ref.~\cite{Kimura:2017gcy}, 
the existence of these primary constraints  (\ref{PRI1}) and (\ref{PRI2})
based on the maximally degenerate conditions is enough to 
remove the extra d.o.f. in the fermionic sector.
However, if these primary constraints yield secondary constraints, we have an even smaller number of physical d.o.f.  Here, let us consider that we have the maximum number of physical d.o.f. in the maximally degenerate Lagrangian, as we have with usual Weyl fields. 
Then, we need to check that no secondary constraints appear by examining the consistency conditions of the primary constraints.
To this end, we first
define the total Hamiltonian density as 
\ba
{\cal H}_T &=& {\cal H}+ \Phi_{\psi_I^\alpha} \lambda_I^{\alpha} + \Phi_{\psib_I^\alphad} {\bar \lambda}_I^\alphad\,,
\label{THamitonian}
\ea
where the Hamiltonian and the total Hamiltonian are given by 
\ba
H=\int d^3x \, {\cal H}, \qquad
H_T=\int d^3x \, {\cal H}_T .
\ea
Now, we would like to calculate the Poisson bracket, defined as
\ba
&&\{\F (t, \bx), \G(t, \by) \} \nonumber\\
&&~~~~= \int d^3 z \Biggl[
{\delta \F(t, \bx)  \over \delta \phi^a (t, \bz) }{\delta \G(t, \by)  \over \delta \pi_{\phi^a} (t, \bz) }
-{\delta \F(t, \bx)  \over  \delta \pi_{\phi^a} (t, \bz) }{\delta \G(t, \by)  \over \delta \phi^a (t, \bz)}\nonumber\\
&&~~~~~~~~~
+(-)^{\varepsilon_\F} 
\left(
{\delta \F(t, \bx)  \over \delta \psi_I^\alpha (t, \bz) }{\delta \G(t, \by)  \over \delta \pi_{\psi_I^\alpha} (t, \bz) }
+{\delta \F(t, \bx)  \over \delta \pi_{\psi_I^\alpha}(t, \bz) }{\delta \G(t, \by)  \over \delta  \psi_I^\alpha (\bz) }
+{\delta \F(t, \bx)  \over \delta \psib_I^\alphad (t, \bz) }{\delta \G(t, \by)  \over \delta \pi_{\psib_I^\alphad} (t, \bz) }
+{\delta \F(t, \bx)  \over \delta  \pi_{\psib_I^\alphad} (t, \bz) }{\delta \G(t, \by)  \over \delta\psib_I^\alphad (t, \bz)}
\right)
\Biggr].\nonumber\\
\ea
The Poisson brackets between the canonical variables are given by
\ba
\{\phi^a (t, \bx), \,\pi_{\phi^b}  (t, \by) \} &=& \delta^a_{~b} \delta^3(\bx -\by) \,,\\
\{\psi_I^\alpha (t, \bx), \, \pi_{\psi_J^\beta}  (t, \by) \} &=& -\delta^{\alpha}_{~\beta}\,\delta_{IJ}\, \delta^3(\bx -\by) \,,\\
\{\psib_I^\alphad (t, \bx), \,\pi_{\psib_J^\betad}  (t, \by)\} &=& -\delta^{\alphad}_{~\betad}\,\delta_{IJ}\, \delta^3(\bx -\by)\,,
\ea
while other Poisson brackets are zero. 
Then, the Poisson brackets between the primary constraints are 
\ba
\{\Phi_{\psi_I^\alpha}(t, \bx), \,\Phi_{\psi_J^\beta}(t, \by) \} &=& 
{\delta F_{\alpha,I} (t, \bx)  \over \delta \psi_J^\beta (t, \by) }
+{\delta F_{\beta,J} (t, \by)  \over \delta  \psi_I^\alpha (t, \bx) }
+\int d^3 z \Biggl[
{\delta F_{\alpha,I}(t, \bx)  \over \delta \phi^a (t, \bz) }{\delta F_{\beta,J}(t, \by)  \over \delta \pi_{\phi^a} (t, \bz) }
-{\delta F_{\alpha,I}(t, \bx)  \over  \delta \pi_{\phi^a} (t, \bz) }{\delta F_{\beta,J}(t, \by)  \over \delta \phi^a (t, \bz)}\biggr]\,, ~~~~~\label{PBPhiPhi1}\\
\{\Phi_{\psi_I^\alpha}(t, \bx), \,\Phi_{\psib_J^\alphad}(t, \by)  \} &=&  
{\delta F_{\alpha, I} (t, \bx)  \over \delta \psib_J^\alphad (t, \by) }
+{\delta G_{\alphad, J} (t, \by)  \over \delta  \psi_I^\alpha (t, \bx) }
+\int d^3 z \Biggl[
{\delta F_{\alpha, I}(t, \bx)  \over \delta \phi^a (t, \bz) }{\delta G_{\alphad, J} (t, \by)  \over \delta \pi_{\phi^a} (t, \bz) }
-{\delta F_{\alpha, I}(t, \bx)  \over  \delta \pi_{\phi^a} (t, \bz) }{\delta G_{\alphad,J}(t, \by)  \over \delta \phi^a (t, \bz)}\biggr]\,, ~~~~~\label{PBPhiPhi2}\\
\{\Phi_{\psib_I^\alphad}(t, \bx), \,\Phi_{\psib_J^\betad}(t, \by) \} &=& 
{\delta G_{\alphad,I} (t, \bx)  \over \delta \psib_J^\betad (t, \by) }
+{\delta G_{\betad, J} (t, \by)  \over \delta  \psib_I^\alphad (t, \bx) }
+\int d^3 z \Biggl[
{\delta G_{\alphad,I}(t, \bx)  \over \delta \phi^a (t, \bz) }{\delta G_{\betad, J}(t, \by)  \over \delta \pi_{\phi^a} (t, \bz) }
-{\delta G_{\alphad,I}(t, \bx)  \over  \delta \pi_{\phi^a} (t, \bz) }{\delta G_{\betad, J}(t, \by)  \over \delta \phi^a (t, \bz)}\biggr]\,. ~~~~~\label{PBPhiPhi3}
\ea
Then, the time evolution of the primary constraints are 
\ba
\begin{pmatrix}
	{\dot \Phi}_{\psi_I^\alpha}(t, \bx) \\
	{\dot \Phi}_{\psib_I^\alphad} (t, \bx)
\end{pmatrix}
=
\left(
\begin{array}{c}
	\{ \Phi_{\psi_I^\alpha} (t, \bx), \, {H}_T\}   \\
	\{ \Phi_{\psib_I^\alphad}(t, \bx), \, {H}_T \}   \\
\end{array}
\right)
=
\left(
\begin{array}{c}
	\{ \Phi_{\psi_I^\alpha} (t, \bx), \, {H}\}   \\
	\{ \Phi_{\psib_I^\alphad}(t, \bx), \, {H} \}   \\
\end{array}
\right)
+
\int d^3 {\bf y} \,
{\bf C}_{IJ}(t, \bx, \by)
\left(
\begin{array}{c}
	\lambda_J^\beta(t, \by)   \\
	{\bar \lambda}_J^\betad(t, \by)   \\
\end{array}
\right)
\approx 0 \,,
\label{TEequation}
\ea
where 
\ba
{\bf C}_{IJ}(t,\bx, \by)=\left(
\begin{array}{cc}
	\{ \Phi_{\psi_I^\alpha}(t, \bx), \,   \Phi_{\psi_J^\beta}(t, \by)\} & \{ \Phi_{\psi_I^\alpha}(t, \bx), \,   \Phi_{\psib_J^\betad}(t, \by)\}\\
	\{ \Phi_{\psib_I^\alphad}(t, \bx), \,   \Phi_{\psi_J^\beta}(t, \by)\} & \{ \Phi_{\psib_I^\alphad}(t, \bx), \,   \Phi_{\psib_J^\betad}(t, \by)\}\\
\end{array}
\right)\,.
\label{MatrixC}
\ea
In order not to have secondary constraints, all the Lagrange multipliers should be fixed by 
the equations~(\ref{TEequation}).  This can be realized if 
the coefficient matrix of the Lagrange multipliers~\eqref{MatrixC} has 
the inverse after integrating over $\by$, and then the primary constraints  (\ref{PRI1}) and (\ref{PRI2}) are second class. In this case, 
the number of d.o.f. is 
\ba
{\textrm{d.o.f.} }= {2 \times n \,\textrm{(boson)} + 2 \times 4N\,\textrm{(fermions)} - 4 N \,\textrm{(constraints)} \over 2} 
= n\,\textrm{(boson)}+2N \,\textrm{(fermions)} \,,
\label{DOFs}
\ea
as desired. 
To find explicit expressions of these obtained conditions, one needs a concrete Lagrangian form. Hereafter, we will consider the most simplest case $N=1$. The extension to multiple Weyl fields can be done in the same way although the analysis will be tedious. A brief introduction for constructing Lagrangians for arbitrary $N$ is given in Appendix~\ref{app:C}.

%%%%%%%%%%%%%%%%%%%%%%%%%%%%%%%%%%%%%
\section{Degenerate scalar-fermion theories}
\label{sec:III}
%%%%%%%%%%%%%%%%%%%%%%%%%%%%%%%%%%%%%
Since the fermionic fields obey the Grassmann algebra, one can dramatically simplify 
the analysis by restricting the number of fermionic fields. Hereafter, we confine the theory to $n$ scalar fields and one Weyl field to simplify the analysis. 
In this section,  we construct the most general scalar-fermion theory of which the Lagrangian 
contains up to quadratic in first derivatives of scalar and fermionic fields.
To this end, we first construct Lorentz scalars and vectors without any derivatives, which consist of the scalar fields $\phi^a$ and the fermionic fields $\psi^\alpha$, $\psib^\alphad$. The fermionic indices can be contracted with the building block matrices, $\eps_{\alpha\beta}$, $\eps_{\alphad\betad}$, $ \sigma_{\alpha\alphad}^\mu$, $(\sigma^{\mu\nu}\varepsilon)_{\alpha\beta}$, and $(\varepsilon\bar{\sigma}^{\mu\nu})_{\alphad\betad}$,\footnote{See Appendix~\ref{app:A} for the detailed definition of each matrix.} and we have three possibilities,
\ba
&& 
\Psi = \psi^\alpha \psi_\alpha , \qquad  
{\bar \Psi} = \psib_\alphad \psib^\alphad, \qquad
J^\mu =\psib^\alphad \sigma_{\alpha\alphad}^\mu \psi^\alpha\,,
\ea
where 
$\Psi^\dagger={\bar \Psi}$. 
Note that any contractions of $(\sigma^{\mu\nu}\varepsilon)_{\alpha\beta}$ and $(\varepsilon\bar{\sigma}^{\mu\nu})_{\alphad\betad}$ with the fermionic fields
always vanish due to the Grassmann properties.\footnote{For $N\neq1$, 
$(\sigma^{\mu\nu}\varepsilon)_{\alpha\beta}$ and $(\varepsilon\bar{\sigma}^{\mu\nu})_{\alphad\betad}$
can be contracted with different Weyl fields, and therefore rank-2 tensors exist. See Appendix~\ref{app:C} for the detail.}
One can also construct a scalar quantity from the square of $J^\mu $
contracted with the Minkowski metric. However, 
it reduces to 
\ba
\eta_{\mu\nu}J^\mu J^\nu = -2\Psi{\bar \Psi},
\ea
where we have used (\ref{B3}). 
Furthermore, the square of $\Psi$ vanishes because of the fact that $\Psi^2 \propto \psi^1 \psi^2 \psi^1 \psi^2$ and the Grassmann property, i.e., 
$\Psi^2 = {\bar \Psi}^2 =0$, while $\Psi {\bar \Psi}$ is nonzero. This implies that 
an arbitrary function ${\cal A} (\phi^a, \Psi, {\bar \Psi}) $ can be expanded in terms of $\Psi$ and ${\bar \Psi}$ as 
\ba
{\cal A} (\phi^a, \Psi, {\bar \Psi}) &=& a_0 (\phi^a) + a_1 (\phi^a) \Psi 
+a_2 (\phi^a) {\bar \Psi}+a_3 (\phi^a) \Psi {\bar \Psi}\,.
\label{arbEX}
\ea
When the arbitrary function $\cal A$ is real, i.e., ${\cal A}={\cal A}^\dagger$, 
then $a_0$ and $a_3$ are also real, and $a_1^\ast=a_2$. 
The inverse of ${\cal A} (\phi^a, \Psi, {\bar \Psi}) $ is given by
\ba
{\cal A}^{-1} (\phi^a, \Psi, {\bar \Psi})
= a_0^{-1} (\phi^a)\left(1- {a_1(\phi^a) \over a_0(\phi^a)}  \Psi -  {a_2(\phi^a) \over a_0(\phi^a)} {\bar \Psi} + {2a_1(\phi^a) a_2(\phi^a) -a_0(\phi^a) a_3(\phi^a) \over a_0(\phi^a)^2}\Psi {\bar \Psi} \right),
\ea
and the condition that ${\cal A} (\phi^a, \Psi, {\bar \Psi})$ has an inverse is simply $a_0(\phi^a) \neq 0$.

As we will show in the following subsection, 
the most general action up to quadratic in first derivatives
of the scalar fields and the Weyl fermion can be written as 
\ba
S=\int d^4 x \, \left( {\cal L}_0 + {\cal L}_1 +{\cal L}_2 \right) \ ,
\label{N=1_action}
\ea
where 
\ba
{\cal L}_0 &=& P_0 \,, \label{L0}\\
{\cal L}_1 &=& 
\PI_a\, \p_\mu \phi^a \, J^\mu
+ \PII \left( \psib^\alphad \sigma_{\alpha\alphad}^\mu\,\p_\mu\psi^\alpha
-\p_\mu \psib^\alphad 
\, \sigma_{\alpha\alphad}^\mu \psi^\alpha \right)
+ \PIII J^\mu \psi_\alpha \p_\mu \psi^\alpha 
+ \PIII{}^\dagger J^\mu \p_\mu \psib^\alphad \psib_\alphad \,,
\label{L1} \\
{\cal L}_2 &=& 
{1 \over 2} V_{ab}^{\mu\nu}\p_\mu \phi^a \,\p_\nu \phi^b
+S_{a\alpha}^{\mu\nu} \p_\mu \phi^a \p_\nu \psi^\alpha 
+ \p_\mu \phi^a \p_\nu \psib^\alphad  \left(S_{a\alpha}^{\mu\nu}\right)^\dagger \nonumber\\
&&
+{1 \over 2}W^{\mu\nu}_{\alpha\beta} \,\p_\mu \psi^\alpha \p_\nu \psi^\beta 
+{1 \over 2}\,\p_\nu \psib^\betad \p_\mu  \psib^\alphad  \left(W^{\mu\nu}_{\alpha\beta} \right)^\dagger
+\p_\mu \psib^\alphad \, Q^{\mu\nu}_{\alpha\alphad} \, \p_\nu \psi^\alpha \label{Lag:L012}\,,
\ea
and
\ba
V_{ab}^{\mu\nu}&=&V_{ab}  \,  \eta^{\mu\nu} \ , \nonumber \\
S^{\mu\nu}_{a\alpha} &=& \SI_a \eta^{\mu\nu} \psi_\alpha + \SII_a (\sigma^{\mu\nu}\eps)_{\alpha\beta} \psi^\beta \nonumber\ , \\
W^{\mu\nu}_{\alpha\beta}&=& 
W_1\, \eta^{\mu\nu} \, \eps_{\alpha\beta}
+W_2 \, \eta^{\mu\nu}  \, \psi_\alpha \psi_\beta 
+W_3 \, (\sigma^{\mu\nu}\eps)_{\alpha\beta} 
+ W_4 \, \Bigl[ \psi_\alpha (\sigma^{\mu\nu}\eps)_{\beta\gamma} + \psi_\beta (\sigma^{\mu\nu}\eps)_{\alpha\gamma} \Bigr]\psi^\gamma \ , \nonumber\\
Q^{\mu\nu}_{\alpha\alphad}&=& 
Q_1 \, \eta^{\mu\nu} \, \psi_\alpha  \psib_\alphad
+Q_2 \,  (\sigma^{\mu\nu}\eps)_{\alpha\beta} \psi^\beta \psib_\alphad
- Q_2^\dagger\, \psi_\alpha \psib^\betad(\eps \sigmab^{\mu\nu})_{\betad\alphad}
+Q_3\,  (\sigma^{\rho\mu}\eps)_{\alpha\beta}  \psi^\beta \psib^\betad (\eps \sigmab^{\rho\nu})_{\betad\alphad}  \label{Lag:VSWQ}\ .
\ea
Here, $P_0$, $P^{(i)}$, $V$, $S^{(i)}$, $W_i$, and $Q_i$ are arbitrary functions of $\phi^a,~\Psi$, and ${\bar \Psi}$.
These arbitrary functions satisfy the following properties : 
\ba
V_{ab}=V_{ba}, \quad
V_{ab} = V_{ab}{}^\dagger, \quad
P_0 = P_0^\dagger, \quad
\PI_a = \PI_a{}^\dagger, \quad
 P^{(2)}=-P^{(2)\dagger} \ , \quad 
Q_1 = Q_1^\dagger, \quad
Q_3 = Q_3^\dagger \,.
\label{supply_prop}
\ea

\subsection{Construction of Lagrangian up to quadratic in the derivatives}

In this subsection, we derive the most general Lagrangian
(\ref{N=1_action})--(\ref{supply_prop}) one by one.

\subsubsection{No derivative term}

The Lagrangian without any derivatives should be the arbitrary function of the form in (\ref{arbEX}); thus,
\ba
{\cal L}_0 &=& P_0(\phi^a, \Psi, \bar \Psi) \,,
\ea
where $P_0$ is an arbitrary function of $\phi^a, \Psi,$ and $\bar \Psi$, and the Hermitian property of the Lagrangian requires $P_0 = P_0^\dagger$.

\subsubsection{Linear terms in the derivatives}
\label{subsec:linear}
The candidate for the Lagrangian including terms proportional to the first derivatives is
\ba
{\cal L}_1 &=& 
\PI_a\, \p_\mu \phi^a \, J^\mu
+ \PII \,  \psib^\alphad \sigma_{\alpha\alphad}^\mu\,\p_\mu\psi^\alpha
+ \PII{}^\dagger \,\p_\mu \psib^\alphad 
\, \sigma_{\alpha\alphad}^\mu \psi^\alpha 
+ \PIII J^\mu \psi_\alpha \p_\mu \psi^\alpha 
+ \PIII{}^\dagger J^\mu \p_\mu \psib^\alphad \psib_\alphad \,,
\ea
where $P^{(i)}$ are arbitrary functions of $\phi^a, \Psi$, and $\bar \Psi$, and the Hermitian property of the Lagrangian requires $\PI_a = \PI_a{}^\dagger$. The second and third terms can be decomposed as
\begin{align}
\PII \,  \psib^\alphad \sigma_{\alpha\alphad}^\mu\,\p_\mu\psi^\alpha
+ \PII{}^\dagger \,\p_\mu \psib^\alphad \, \sigma_{\alpha\alphad}^\mu \psi^\alpha 
= \frac{1}{2}(\PII + \PII{}^\dagger) \partial_\mu J^\mu +
\frac{1}{2}(\PII - \PII{}^\dagger) (\psib^\alphad \sigma_{\alpha\alphad}^\mu\,\p_\mu\psi^\alpha -\p_\mu \psib^\alphad \, \sigma_{\alpha\alphad}^\mu \psi^\alpha ) \ .
\end{align}
If we integrate the first term in the right-hand side by parts, it becomes
\ba
-\frac{1}{2}(\PII + \PII{}^\dagger)_{, \phi^a}\p_\mu \phi^a \, J^\mu
+(\PII + \PII{}^\dagger)_{, \Psi}J^\mu \psi_\alpha \p_\mu \psi^\alpha 
+(\PII + \PII{}^\dagger)_{, \Psib}J^\mu \p_\mu \psib^\alphad \psib_\alphad 
\ea
where $P_{,z} \equiv \p P / \p z$ and $P^\dagger_{,z} \equiv \p P^\dagger / \p z$.
Thus, the real part of $\PII$ can be absorbed into the other terms in ${\cal L}_1$, 
and thus we can impose $\PII=-\PII{}^\dagger$ without loss of generality.

\subsubsection{Quadratic terms in the derivatives}
Next, the Lagrangian containing two derivatives is 
\ba
{\cal L}_2 &=&
{1 \over 2}V_{ab}^{\mu\nu}\p_\mu \phi^a \,\p_\nu \phi^b
+S_{a\alpha}^{\mu\nu} \p_\mu \phi^a \p_\nu \psi^\alpha + \p_\mu \phi^a \p_\nu \psib^\alphad (S^{\mu\nu}_{a\alpha})^\dagger \nonumber\\
&&+{1 \over 2}W^{\mu\nu}_{\alpha\beta} \,\p_\mu \psi^\alpha \p_\nu
\psi^\beta +{1 \over 2} \p_\nu \psib^\betad \p_\mu \psib^\alphad \, (W^{\mu\nu}_{\alpha\beta})^\dagger \,
+\p_\mu \psib^\alphad \, Q^{\mu\nu}_{\alpha\alphad} \, \p_\nu \psi^\alpha \,,
\ea
where $V^{\mu\nu}_{ab}$, $S_{a\alpha}^{\mu\nu}$, $W^{\mu\nu}_{\alpha\beta}$, and $Q^{\mu\nu}_{\alpha\alphad}$
consist of $\phi^a$, $\psi^\alpha$, $\psib^\alphad$, and constant matrices. 
The coefficient $V^{\mu\nu}_{ab}$ should take the form 
\ba
V^{\mu\nu}_{ab}=V_{ab}^{(1)}  \,  \eta^{\mu\nu} + V_{ab}^{(2)}  \,  J^\mu J^\nu, 
\ea
where $V_{ab}^{(i)}$ are arbitrary functions of $\phi^a, \Psi,$ and $\bar \Psi$. 
However, the identity \eqref{B8} leads to 
\ba
J^\mu J^\nu = -{1\over 2} \eta^{\mu\nu} \Psi {\bar \Psi}.
\label{JJ}
\ea
Therefore, the $V_{ab}^{(2)}$ term can be absorbed into the $V_{ab}^{(1)}$ term, and we can generally set $V_{ab}^{(2)}=0$.
As a result, we have
\begin{align}
 V^{\mu\nu}_{ab}=V_{ab}  \,  \eta^{\mu\nu} \ ,
\end{align}
where $V_{ab}$ is an arbitrary real function of $\phi^a, \Psi,$ and $\bar \Psi$, 
which is symmetric under the exchange of $a$ and $b$.

Second, the general form of $S^{\mu\nu}_{a\alpha}$ is 
\ba
S^{\mu\nu}_{a\alpha} = \SI_a \eta^{\mu\nu} \psi_\alpha 
+ \SII_a J^\mu \psib^\alphad \sigma_{\alpha\alphad}^\nu 
+ \SIII_a J^\nu \psib^\alphad \sigma_{\alpha\alphad}^\mu 
+ \SIIII_a J^\mu J^\nu \psi_\alpha \,,
\label{coeff_S_gene}
\ea
where $S_a^{(i)}$ are arbitrary functions of $\phi^a, \Psi,$ and $\bar \Psi$. 
By using (\ref{B7}) and (\ref{B8}), 
the second and third terms can be rewritten as 
\ba
J^\mu \psib^\alphad \sigma_{\alpha\alphad}^\nu &=& {1\over 2} {\bar \Psi} \eta^{\mu\nu} \psi_\alpha +  {\bar \Psi}(\sigma^{\mu\nu}\eps)_{\alpha\beta} \psi^\beta \,,
\label{id_S_aalpha1}\\
J^\nu \psib^\alphad \sigma_{\alpha\alphad}^\mu &=&{1\over 2} {\bar \Psi} \eta^{\mu\nu} \psi_\alpha -  {\bar \Psi}(\sigma^{\mu\nu}\eps)_{\alpha\beta} \psi^\beta \, . \label{id_S_aalpha2}
\ea
Therefore, the second and third terms can be compactly expressed as $\SII_a (\sigma^{\mu\nu}\eps)_{\alpha\beta} \psi^\beta$, 
which is antisymmetric under $\mu$ and $\nu$,
by absorbing the symmetric remaining parts into $\SI_a$. 
The last term with $\SIIII_a$ is proportional to $\psi^1\psi^1$ or $\psi^2 \psi^2$ in a component expression;
thus, this automatically vanishes. 
As a result, we can assign
\begin{align}
 S^{\mu\nu}_{a\alpha} = \SI_a \eta^{\mu\nu} \psi_\alpha + \SII_a (\sigma^{\mu\nu}\eps)_{\alpha\beta} \psi^\beta  \ .
\end{align}

Third, the general form of $W^{\mu\nu}_{\alpha\beta}$ is 
\ba
W^{\mu\nu}_{\alpha\beta}&=& 
W_1\, \eta^{\mu\nu} \, \eps_{\alpha\beta}
+W_2 \, \eta^{\mu\nu}  \, \psi_\alpha \psi_\beta 
+ \widetilde{W}_3\,  (\psib^\alphad \sigma_{\alpha \alphad}^\mu) (\psib^\betad \sigma_{\beta \betad}^\nu) 
+ \widetilde{W}_4\,  (\psib^\alphad \sigma_{\alpha \alphad}^\nu) (\psib^\betad \sigma_{\beta \betad}^\mu) \nonumber\\
&&+ W_5\,  J^\mu \psi_\alpha \psib^\betad \sigma_{\beta\betad}^\nu
+ W_6\, J^\nu \psi_\alpha \psib^\betad \sigma_{\beta\betad}^\mu
+W_7\, J^\mu J^\nu  \, \psi_\alpha \psi_\beta \,.
\ea
We note that $J^\mu\psi_\beta\bar{\psi}^\betad\sigma^\nu_{\alpha\betad}\simeq -J^\nu\psi_\alpha\bar{\psi}^\betad\sigma^\mu_{\beta\betad}$ and $J^\nu\psi_\beta\bar{\psi}^\betad\sigma^\mu_{\alpha\betad}\simeq -J^\mu\psi_\alpha\bar{\psi}^\betad\sigma^\nu_{\beta\betad}$ hold as long as they are contracted with $\partial_\mu\psi^\alpha\partial_\nu\psi^\beta$. Similarly to the case of $S^{(2)}$ and $S^{(3)}$, we can rewrite $\widetilde{W}_3$, $\widetilde{W}_4$, $W_5$, and $W_6$ as 
\ba
(\psib^\alphad \sigma_{\alpha \alphad}^\mu) (\psib^\betad \sigma_{\beta \betad}^\nu)  &=& {1\over 2} {\bar \Psi} \eta^{\mu\nu} \eps_{\alpha\beta} -  {\bar \Psi}(\sigma^{\mu\nu}\eps)_{\alpha\beta}  \,,\\
(\psib^\alphad \sigma_{\alpha \alphad}^\nu) (\psib^\betad \sigma_{\beta \betad}^\mu) &=& {1\over 2} {\bar \Psi} \eta^{\mu\nu} \eps_{\alpha\beta} +  {\bar \Psi}(\sigma^{\mu\nu}\eps)_{\alpha\beta}  \, .\\
 J^\mu \psi_\alpha \psib^\betad \sigma_{\beta\betad}^\nu  &=& {1\over 2} {\bar \Psi} \eta^{\mu\nu} \psi_\alpha \psi_\beta +  {\bar \Psi} \psi_\alpha (\sigma^{\mu\nu}\eps)_{\beta\gamma} \psi^\gamma \,,\\
J^\nu \psi_\alpha \psib^\betad \sigma_{\beta\betad}^\mu &=& {1\over 2} {\bar \Psi} \eta^{\mu\nu} \psi_\alpha \psi_\beta -  {\bar \Psi} \psi_\alpha (\sigma^{\mu\nu}\eps)_{\beta\gamma} \psi^\gamma  \, .
\ea
Thus, the symmetric parts of $\widetilde{W}_3$, $\widetilde{W}_4$ and $W_5$, $W_6$ can be expressed in terms of $\eta^{\mu\nu}\varepsilon_{\alpha\beta}$ and $\eta^{\mu\nu}\psi_{\alpha}\psi_{\beta}$. On the other hand, antisymmetric parts of them are taken into account by $(\sigma^{\mu\nu}\eps)_{\alpha\beta}$ and $\psi_\alpha (\sigma^{\mu\nu}\eps)_{\beta\gamma} \psi^\gamma$. 
As before, the last term with $W_7$ is proportional to $\psi^1\psi^1$ in a component expression; thus, this automatically vanishes. 
As a result, we can express $W^{\mu\nu}_{\alpha\beta}$ as
\begin{align}
 W^{\mu\nu}_{\alpha\beta}=
W_1\, \eta^{\mu\nu} \, \eps_{\alpha\beta}
+W_2 \, \eta^{\mu\nu}  \, \psi_\alpha \psi_\beta 
+W_3 \, (\sigma^{\mu\nu}\eps)_{\alpha\beta} 
+ W_4 \, \Bigl[ \psi_\alpha (\sigma^{\mu\nu}\eps)_{\beta\gamma} + \psi_\beta (\sigma^{\mu\nu}\eps)_{\alpha\gamma} \Bigr]\psi^\gamma \ .
\end{align}
The $W_4$ term is manifestly symmetrized by making use of the fact that $\psi_\alpha(\sigma^{\mu\nu}\varepsilon)_{\beta\gamma}\psi^\gamma\simeq\psi_\beta(\sigma^{\mu\nu}\varepsilon)_{\alpha\gamma}\psi^\gamma$ in the Lagrangian, which holds again since they are contracted with  $\partial_\mu \psi^\alpha \partial_\nu \psi^\beta$.

Finally, let us take a look at $Q^{\mu\nu}_{\alpha\alphad}$. The general form is given by
\ba
Q^{\mu\nu}_{\alpha\alphad}&=& 
Q_1 \, \eta^{\mu\nu} \, \psi_\alpha  \psib_\alphad
+\widetilde{Q}_2 \, (\psib^\betad \sigma_{\alpha\betad}^\mu) (\sigma_{\beta\alphad}^\nu \psi^\beta)
+ \widetilde{Q}_3\, (\psib^\betad \sigma_{\alpha\betad}^\nu) (\sigma_{\beta\alphad}^\mu \psi^\beta)
+Q_4\, J^\mu \sigma_{\alpha\alphad}^\nu
+Q_5\, J^\nu \sigma_{\alpha\alphad}^\mu
\nonumber\\
&&+Q_6 \, J^\mu J^\nu \,  \psi_\alpha  \psib_\alphad
+Q_7 \,  J^\mu \psi_\alpha \psi^\beta \sigma_{\beta\alphad}^\nu
+ Q_8\,    J^\nu \psi_\alpha \psi^\beta \sigma_{\beta\alphad}^\mu\, 
+ Q_9 \, J^\mu \psib_\alphad \psib^\betad \sigma_{\alpha\betad}^\nu
+Q_{10} \, J^\nu \psib_\alphad \psib^\betad \sigma^\mu_{\alpha\betad}
\ea
The last five terms, $Q_{6, 7, 8, 9, 10}$, contain more than one $\psi^1$, $\psi^2$,  $\bar{\psi}^{\dot{1}}$ or $\psib^{\dot{2}}$, and therefore they automatically vanish.
We can rewrite $\widetilde{Q}_2$, $\widetilde{Q}_3$, $Q_4$, and $Q_5$ as 
\ba
(\psib^\betad \sigma_{\alpha\betad}^\mu) (\sigma_{\beta\alphad}^\nu \psi^\beta) &=& 
-{1\over 2}  \eta^{\mu\nu}  \psi_\alpha \psib_\alphad
+ (\sigma^{\mu\nu}\eps)_{\alpha\beta} \psi^\beta  \psib_\alphad
-  \psi_\alpha (\eps \sigmab^{\mu\nu})_{\alphad\betad}\psib^\betad
-2 (\sigma^{\rho\mu}\eps)_{\alpha\beta}  \psi^\beta (\eps \sigmab^{\rho\nu})_{\alphad\betad} \psib^\betad 
\,,\\
(\psib^\betad \sigma_{\alpha\betad}^\nu) (\sigma_{\beta\alphad}^\mu \psi^\beta) &=& 
-{1\over 2}  \eta^{\mu\nu}  \psi_\alpha \psib_\alphad
- (\sigma^{\mu\nu}\eps)_{\alpha\beta} \psi^\beta  \psib_\alphad
+  \psi_\alpha (\eps \sigmab^{\mu\nu})_{\alphad\betad}\psib^\betad
-2 (\sigma^{\rho\mu}\eps)_{\alpha\beta}  \psi^\beta (\eps \sigmab^{\rho\nu})_{\alphad\betad} \psib^\betad 
 \, .\\
 J^\mu \sigma_{\alpha\alphad}^\nu  &=& 
{1\over 2}  \eta^{\mu\nu}  \psi_\alpha \psib_\alphad
+ (\sigma^{\mu\nu}\eps)_{\alpha\beta} \psi^\beta  \psib_\alphad
+  \psi_\alpha (\eps \sigmab^{\mu\nu})_{\alphad\betad}\psib^\betad
-2 (\sigma^{\rho\mu}\eps)_{\alpha\beta}  \psi^\beta (\eps \sigmab^{\rho\nu})_{\alphad\betad} \psib^\betad 
 \,,\\
J^\nu \sigma_{\alpha\alphad}^\mu&=&
{1\over 2}  \eta^{\mu\nu}  \psi_\alpha \psib_\alphad
- (\sigma^{\mu\nu}\eps)_{\alpha\beta} \psi^\beta  \psib_\alphad
-  \psi_\alpha (\eps \sigmab^{\mu\nu})_{\alphad\betad}\psib^\betad
-2 (\sigma^{\rho\mu}\eps)_{\alpha\beta}  \psi^\beta (\eps \sigmab^{\rho\nu})_{\alphad\betad} \psib^\betad 
  \, .
\ea
Note that the last terms are symmetric under the replacement of $\mu$ and $\nu$, as shown in \eqref{B10}.
Thus, four functions from $Q_{1,4,5}$ and $\widetilde{Q}_{2,3}$ are independent.
We adopt
$\eta^{\mu\nu}  \psi_\alpha \psib_\alphad,\, (\sigma^{\mu\nu}\eps)_{\alpha\beta} \psi^\beta  \psib_\alphad,\, \psi_\alpha (\eps \sigmab^{\mu\nu})_{\alphad\betad}\psib^\betad$, and $(\sigma^{\rho\mu}\eps)_{\alpha\beta}  \psi^\beta (\eps \sigmab^{\rho\nu})_{\alphad\betad} \psib^\betad $ as independent functions, and then we have
\begin{align}
 Q^{\mu\nu}_{\alpha\alphad}=
Q_1 \, \eta^{\mu\nu} \, \psi_\alpha  \psib_\alphad
+Q_2 \,  (\sigma^{\mu\nu}\eps)_{\alpha\beta} \psi^\beta \psib_\alphad
- Q_2^\dagger\, \psi_\alpha \psib^\betad(\eps \sigmab^{\mu\nu})_{\betad\alphad}
+Q_3\,  (\sigma^{\rho\mu}\eps)_{\alpha\beta}  \psi^\beta \psib^\betad (\eps \sigmab^{\rho\nu})_{\betad\alphad} \ .
\end{align}
Since the first and the fourth terms are Hermite, the coefficients are required to be real.

Collecting all the results, we obtain
the most general Lagrangian (\ref{N=1_action})--(\ref{supply_prop}).

\subsection{Degeneracy conditions}
In the previous subsection, we have obtained the most general Lagrangian of $n$-scalar and one Weyl fields which contains up to quadratic in first derivatives.
As explained earlier, the Euler-Lagrange equations, in general, contain second derivatives of the fermionic fields. Thus, one should choose the arbitrary functions appearing 
in the Lagrangian with care in order to have the correct number of d.o.f. 
In this subsection, we derive the degeneracy conditions for the Lagrangian 
\eqref{Lag:L012} with \eqref{Lag:VSWQ}. (Note that the linear Lagrangian in derivatives~\eqref{L1} is irrelevant to the degeneracy conditions.)
The maximally degenerate conditions \eqref{MDC1} and \eqref{MDC2}  lead to 
\begin{align}
 D_{\alpha\beta}-\C_{\alpha b} A^{ba} \B_{a\beta}
&=W_1 \eps_{\alpha\beta} + \left(W_2 + \SI_b V^{ba} \SI_a \right)\psi_\alpha \psi_\beta=0\,,  \label{MDC1_single}\\
D_{\alpha\betad}-\C_{\alpha b} A^{ba} \B_{a\betad}
&=-\left(Q_1+\SI_b V^{ba} (\SI_a)^\dagger\right)\psi_\alpha\psib_\betad - Q_3 (\sigma^{i0}\eps)_{\alpha\gamma} \psi^\gamma \psib^\gammad (\eps \sigmab^{i0})_{\gammad\dot{\beta}}=0 \ ,\label{MDC2_single}
\end{align}
where $V^{ab}=(V^{-1})^{ab}$. In general, they give us four
conditions,\footnote{As far as we seek a healthy theory where we have
no ghost not only in background but also in the perturbations, these
four conditions are reasonable; e.g., even if the background evolution
is like $\psi_\alpha \psi_{\beta} \propto \varepsilon_{\alpha\beta}$ as a
result of the equations of motion, the perturbations on it will face serious
instabilities.}
\begin{align}
W_1 = 0, \qquad
(W_2 + \SI_b V^{ba} \SI_a ) \psi_\alpha\psi_\beta= 0, \qquad
(Q_1+\SI_b V^{ba} (\SI_a)^\dagger)\psi_\alpha\psib_\betad=0, \qquad
Q_3\psi^\gamma\bar{\psi}^{\dot{\gamma}}=0 \,.
\end{align}
The functions included in ${\cal L}_2$ are simplified after inserting
these four conditions:
\ba
V_{ab}^{\mu\nu}&=&V_{ab}  \,  \eta^{\mu\nu} \ , \nonumber \\
S^{\mu\nu}_{a\alpha} &=& \SI_a \eta^{\mu\nu} \psi_\alpha + \SII_a (\sigma^{\mu\nu}\eps)_{\alpha\beta} \psi^\beta \nonumber\ , \\
W^{\mu\nu}_{\alpha\beta}&=& 
-S_b^{(1)}V^{ba}S_a^{(1)} \, \eta^{\mu\nu}  \, \psi_\alpha \psi_\beta 
+W_3 \, (\sigma^{\mu\nu}\eps)_{\alpha\beta} 
+ W_4 \, \Bigl[ \psi_\alpha (\sigma^{\mu\nu}\eps)_{\beta\gamma} + \psi_\beta (\sigma^{\mu\nu}\eps)_{\alpha\gamma} \Bigr]\psi^\gamma \ , \nonumber\\
Q^{\mu\nu}_{\alpha\alphad}&=& 
-S_b^{(1)}V^{ba}(S_a^{(1)})^\dagger \, \eta^{\mu\nu} \, \psi_\alpha  \psib_\alphad
+Q_2 \,  (\sigma^{\mu\nu}\eps)_{\alpha\beta} \psi^\beta \psib_\alphad
- Q_2^\dagger\, \psi_\alpha \psib^\betad(\eps \sigmab^{\mu\nu})_{\betad\alphad} \ .
\label{max_dege_coeff}
\ea

\section{Hamiltonian formulation}
\label{sec:IV}

In the previous section, we derived the Lagrangian satisfying
the degeneracy conditions which lead to four primary constraints, as
 explicitly shown in the following. They suggest that the extra d.o.f. associated with fermionic Ostrogradsky's ghosts are removed; however, one needs to make sure of the condition that no more (secondary) constraints arise in order to have four physical d.o.f. in the phase space. Following
Sec.~\ref{sec:II}, we derive such a condition. 

The momenta can be directly calculated from their definition,
\ba
\pi_{\phi^a} &=& \PI_a J^0 + V_{ab} \phid^b + S_{a\alpha}^{0\nu} \p_\nu \psi^\alpha
+ \p_\nu \psib^\alphad (S^{0\nu}_{a\alpha})^\dagger  \label{Pi_phi_1} \ ,\\
\pi_{\psi^\alpha} &=&  -\PII \psib^\alphad \sigma_{\alpha\alphad}^0 -\PIII J^0 \psi_\alpha
-S^{\nu0}_{a\alpha} \p_\nu \phi^a + W^{0\nu}_{\alpha\beta} \p_\nu \psi^\beta 
-\p_\nu \psib{}^\betad Q_{\alpha\betad}^{\nu0} \label{Pi_psi_1} \ ,\\
\pi_{\psib^\alphad} &=&- (\pi_{\psi^\alpha} )^\dagger\label{Pi_psid_1} \,.
\ea
Solving \eqref{Pi_phi_1} for $\phid^a$ and plugging it into \eqref{Pi_psi_1} and \eqref{Pi_psid_1}, 
we indeed obtain primary constraints (\ref{PRI1}) and (\ref{PRI2}),
where $F_\alpha$ and $G_\alphad$ are given by
\ba
F_\alpha &=& -\PII \bar{\psi}^\alphad \sigma^0_{\alpha\alphad} -\PIII J^0 \psi_\alpha
-S^{00}_{a\alpha} V^{ab} \left[\pi_{\phi^b} - \PI_b J^0 - S^{0i}_{b\beta} \p_i \psi^\beta - \p_i \psib^\betad (S^{0i}_{b\beta})^\dagger\right] \nonumber\\
&& -S^{i0}_{a\alpha} \p_i\phi^a + W^{0i}_{\alpha\beta} \p_i \psi^\beta - \p_i \psib^\betad Q^{i0}_{\alpha\betad}\,,\\
G_\alphad &=& -(F_\alpha)^\dagger \ .
\ea
Then, plugging these expression into (\ref{PBPhiPhi1}) and (\ref{PBPhiPhi2}) and picking up the bosonic parts, we obtain
\ba
\{\Phi_{\psi^\alpha}(t, \bx), \,\Phi_{\psi^\beta}(t, \by) \}^{(0)}
&=& 2  (\SII_a)^{(0)} \p_i \phi^a (\sigma^{0i} \eps)_{\alpha\beta}\delta(\bx-\by)  \nonumber\\
&&+ (\sigma^{0i}\eps)_{\alpha\beta}
\[
W_3^{(0)} (\bx) {\p \over \p x^i} \delta(\bx-\by) + W_3^{(0)} (\by) {\p \over \p y^i} \delta(\bx-\by)
\]\,,~~\label{PhiPhi}  \\
\{\Phi_{\psi^\alpha}(t, \bx), \,\Phi_{\psib^\betad}(t, \by)  \}^{(0)}
&=& -2 (\PII)^{(0)} \sigma^0_{\alpha\betad}\delta(\bx-\by) \,,
\ea
where $(0)$ represents their bosonic parts.
When we calculate the time derivative of the primary constraints,   
we integrate the product of the constraints matrix and the Lagrange multipliers over $\by$. Focusing on the second term in (\ref{PhiPhi}), 
\ba
&&\int d^3 {\bf y} \,
(\sigma^{0i}\eps)_{\alpha\beta}\[
W_3^{(0)} (\bx) {\p \over \p x^i} \delta(\bx-\by) + W_3^{(0)} (\by) {\p \over \p y^i} \delta(\bx-\by)
\]
\lambda^\beta(t, \by) \nonumber\\
=&&-\int d^3 {\bf y} \,
(\sigma^{0i}\eps)_{\alpha\beta}
 {\p \over \p y^i} \left( W_3^{(0)}(\by) \right)\, \delta(\bx-\by) \lambda^\beta(t, \by)\,
\ea
holds as far as we drop total derivative terms,
where we have used the fact that $(\p/ \p x^i)  \delta(\bx-\by) = -(\p/ \p y^i)  \delta(\bx-\by) $. Using this result and the following identities, 
\begin{align}
 \{\Phi_{\psib^\alphad}(t, \bx), \,\Phi_{\psib^\betad}(t, \by) \} &=& -\{\Phi_{\psi^\alpha}(t, \bx), \,\Phi_{\psi^\beta}(t, \by) \}^\dagger \ ,\quad
 \{\Phi_{\psib^\alphad}(t, \bx), \,\Phi_{\psi^\beta}(t, \by)  \}
=-\{\Phi_{\psi^\alpha}(t, \bx), \,\Phi_{\psib^\betad}(t, \by)  \}^\dagger \ ,
\end{align}
we can write the bosonic part of the constraint matrix \eqref{MatrixC} as 
\begin{align}
{\bf C}^{(0)}(t, \bx, \by)=
\begin{pmatrix}
 (2S_a^{(2)}\partial_i\phi^a- \partial_i W_3)(\sigma^{0i}\varepsilon)_{\alpha\beta} 
 &-2P^{(2)}\sigma^0_{\alpha\betad}\\
 -2P^{(2)}\sigma^0_{\beta\alphad}
 & -(2S_a^{(2)\dagger }\partial_i\phi^a- \partial_i W_3^\dagger)(\varepsilon\bar{\sigma}^{0i})_{\alphad\betad}
\end{pmatrix}^{(0)}
\delta(\bx-\by) \ . 
\label{C0}
\end{align}
After performing the integration in \eqref{TEequation}, we can define the matrix ${\bf J}_H^{(0)} (t, \bx)$,  thanks to the delta function,　as 
\ba
{\bf J}_H^{(0)} (t, \bx) =  \int d^3 {\bf y} \, {\bf C}^{(0)}(t, \bx, \by) \ .
\label{JH0}
\ea
If this matrix ${\bf J}_H^{(0)} (t, \bx) $ is invertible, i.e., 
the determinant is nonzero,\footnote{Allowing the theory to have solutions with $\phi^a=\phi^a(t)$ requires nonvanishing $P^{(2)}$.}
\ba
\det {\bf J}_H^{(0)} (t, \bx) \neq 0 \,,
\label{con:detJH}
\ea
all the Lagrange multipliers introduced in the total Hamiltonian \eqref{THamitonian} are 
determined by solving the simultaneous equations \eqref{TEequation}. 
In this case, the theory does not have secondary constraints, and 
all the (primary) constraints are second class. Therefore, the number of d.o.f. 
is $n+2$ as counted in \eqref{DOFs}, and the extra d.o.f. are properly removed.

\section{Euler-Lagrange equations}\label{Euler-Lagrange}
\label{sec:V}

In this section, we show that the degeneracy conditions
and the supplementary condition obtained 
in the Hamiltonian formulation can be also derived in the Lagrangian formulation. In addition, 
we show that if the supplementary condition is satisfied, the equations of motion for fermions
can always be solved in terms of the first derivative of fermionic fields.

\subsection{Equations of motion for fermions with the maximally degenerate conditions}

We derive the equations of motion for the fields and see the EOMs for fermions become the first-order (nonlinear) differential equations after applying the maximally degenerate conditions.

The variations with respect to the scalar and Weyl fields
yield a set of  Euler-Lagrange equations, 
\ba
\left(
\begin{array}{ccc}
	V^{00}_{ab}  & S_{a\beta}^{00} & -(S_{a\beta}^{00})^\dagger \\
	-S^{00}_{b\alpha}  & W_{\alpha\beta}^{00} & -Q^{00}_{\alpha\betad}  \\
	(S_{b\alpha}^{00})^\dagger & Q^{00}_{\beta\alphad} & -(W^{00}_{\alpha\beta})^\dagger\\
\end{array}
\right)
\left(
\begin{array}{c}
	 \phidd^b  \\
	\psidd^\beta   \\
	\psibdd{}^\betad  \\
\end{array}
\right)
&=&
\left(
\begin{array}{c}
	E_a \\
	\E_\alpha \\
	\E_\alphad \\
\end{array}
\right) \ ,\label{EoM_matrix}
\ea
where the coefficient matrix in the left-hand side corresponds the kinetic matrix in \eqref{KineticMatrix},
and we have defined
\ba
E_a &=& {\p \Lag \over \p \phi^a} -\p_\mu (\PI_a J^\mu) - \p_\mu V_{ab}^{\mu\nu} \p_\nu \phi^b
-\p_\mu S_{a\alpha}^{\mu\nu} \p_\nu \psi^\alpha -\p_\nu \psib{}^\alphad \p_\mu (S^{\mu\nu}_{a\alpha})^\dagger 
\nonumber\\
&&
 -V_{ab}^{ij} \p_i\p_j \phi^b - S^{ij}_{a\alpha} \p_i\p_j \psi^\alpha - \p_i\p_j \psib{}^\alphad (S^{ij}_{a\alpha})^\dagger \,, \label{Ea_single}\\
\E_\alpha &=&
{\p \Lag \over \p \psi^\alpha} + \p_\mu (\PII \psib^\alphad) \sigma_{\alpha\alphad}^\mu
+\p_\mu (\PIII J^\mu \psi_\alpha) + \p_\mu S^{\nu\mu}_{a\alpha} \p_\nu \phi^a 
-\p_\mu W^{\mu\nu}_{\alpha\beta} \p_\nu \psi^\beta + \p_\nu \psib^\alphad \p_\mu Q^{\nu\mu}_{\alpha\alphad} \nonumber\\
&&+S_{a\alpha}^{ij} \p_i\p_j \phi^a -W^{ij}_{\alpha\beta} \p_i\p_j \psi^\beta 
+2\p_i\psibd{}^\alphad Q^{(0i)}_{\alpha\alphad} 
+ \p_i\p_j \psib{}^\alphad Q^{ij}_{\alpha\alphad} \,,  \label{Ealpha_single}
\ea
where $(0i)$ means symmetrized with respect to $0$ and $i$, and $\E_\alphad = -(\E_\alpha)^\dagger$. 
Solving \eqref{Ea_single} for $\phidd$, we obtain
\ba
\phidd^a = V^{ab}\left[E_b- S_{b\alpha}^{00} \psidd^\alpha - \psibdd{}^\alphad (S_{b\alpha}^{00})^\dagger\right] \label{EQ:phi}\,.
\ea
Eliminating $\phidd$ from the second lines of \eqref{EoM_matrix}, we find the equations of motion for fermions,
\begin{align}
& \bigl[W^{00}_{\alpha\beta}+S^{00}_{b\alpha}V^{ba}S^{00}_{a\beta}\bigr]\ddot{\psi}^\beta-\bigl[Q^{00}_{\alpha\betad}+S^{00}_{b\alpha}V^{ba}(S^{00}_{a\beta})^\dagger\bigr]\ddot{\bar{\psi}}^\betad=\E_\alpha + S_{b\alpha}^{00} V^{ba} E_a \ .
\end{align}
The condition that the dependence of the second derivative of fermions vanishes is exactly the same as the degeneracy conditions~\eqref{MDC1_single}--\eqref{MDC2_single}, and the third line of  \eqref{EoM_matrix} is also reduced to the Hermitian conjugate. 
Imposing the maximally degenerate conditions, we obtain
the first-order differential equations for $\psi^\alpha$ and $\psib^\alphad$,  
\ba
\Y_\alpha &\equiv& \E_\alpha + S_{a\alpha}^{00} V^{ab} E_b=0, \label{Y1}\\
\Y_\alphad &\equiv&  -(\Y_\alpha)^\dagger = \E_\alphad - \left(S_{a\alpha}^{00}\right)^\dagger V^{ab} E_b=0 \label{Y2} ,
\ea
and, by substituting \eqref{Ea_single} and \eqref{Ealpha_single}, $\cal{Y}_\alpha$ is written as
\begin{align}
\cal{Y}_\alpha =&
{\p \Lag \over \p \psi^\alpha} + \p_\mu (\PII \psib^\alphad) \sigma_{\alpha\alphad}^\mu
+\p_\mu (\PIII J^\mu \psi_\alpha) + \p_\mu S^{\nu\mu}_{a\alpha} \p_\nu \phi^a 
-\p_\mu W^{\mu\nu}_{\alpha\beta} \p_\nu \psi^\beta + \p_\nu \psib^\alphad \p_\mu Q^{\nu\mu}_{\alpha\alphad} \nonumber\\
&+S_a^{(1)}V^{ab}\psi_\alpha\Bigl[ {\p \Lag \over \p \phi^b} -\p_\mu (\PI_b J^\mu) - \p_\mu V_{bc} \p^\mu \phi^c-\p_\mu S_{b\beta}^{\mu\nu} \p_\nu \psi^\beta -\p_\nu \psib{}^\betad \p_\mu (S^{\mu\nu}_{b\beta})^\dagger 
\Bigr]\ .  \label{EoM_fermion}
\end{align}
We note that the dependence on $\partial_i \dot{\bar{\psi}}_\alphad$ and the second spatial derivatives of the fermions vanishes after applying \eqref{max_dege_coeff}, and they have become the first-order differential equations for both time and spatial derivatives. 

As we saw, thanks to the maximally degenerate conditions, one can remove the second derivative terms of the Weyl field in the Euler-Lagrange equations by combining the equations of motion for the scalar fields,
suggesting that the number of initial conditions to solve the EOMs is appropriate. Thus, we have confirmed even in the Lagrangian formulation that the extra d.o.f. do not appear in the theory with the maximally 
degenerate conditions. Note that the second derivative terms of the Weyl field
in the equations of motion for the scalar fields \eqref{EQ:phi} can be removed 
by using the time derivative of \eqref{Y1} and \eqref{Y2}, 
as discussed in Ref.~\cite{Kimura:2017gcy}.

%%%%%%%%%%%%%%%%%%%%%%%%%%%%%%%%%%%%%
\subsection{Solvable condition for nonlinear equations including first derivatives}
%%%%%%%%%%%%%%%%%%%%%%%%%%%%%%%%%%%%%

So far, we have seen that the maximally degenerate conditions lead to 
the first- (second-)order differential equations of motion for the Weyl (scalar) fields.
As discussed in Ref.~\cite{Kimura:2017gcy}, it is not clear whether 
the equations of motion for fermions can be solved or not due to 
the Grassmann properties. Here, we derive a condition, under which
the equations of motion even nonlinear in the time derivatives
can be explicitly solved. We start with the simplest example in a point particle 
system, and then we extend this analysis to the theory we focus on in the present paper.

\subsubsection{Example : Two Grassmann-odd variables}

Suppose that we have a system composed of bosons $q^i$ and two fermionic variables $\theta_1$ and $\theta_2$. Due to the maximally degenerate conditions, 
the equations of motion for the fermionic variables should be the first-order differential 
equations after appropriately combining the bosonic equations of motion. 
Let us assume that we have already done this process, and we can generally write down the reduced equations of motion for fermionic variables as
\ba
a_1 \thetad_1  + a_2 \thetad_2 + a_3 \thetad_1 \thetad_2 &=& a_4, \nonumber\\
b_1 \thetad_1  + b_2 \thetad_2 + b_3 \thetad_1 \thetad_2 &=& b_4 , \label{NL_EoM}
\ea
where $a_1$, $a_2$, $b_1$, and $b_2$ ($a_3$, $a_4$, $b_3$, and $b_4$) are Grassmann-even (Grassmann-odd) functions depending on $\theta_\alpha$, $q^i$, and $\dot{q}^i$. 
We first assume at least one of $a_1^{(0)}$, $a_2^{(0)}$, $b_1^{(0)}$, and $b_2^{(0)}$ is nonzero.
For instance, if $a^{(0)}_1\neq 0$, we can solve the first equation of \eqref{NL_EoM} for $\thetad_1$:
\ba
\thetad_1 = a_1^{-1}a_4+a_1^{-1}\bigl(-a_2+a_3 a_1^{-1} a_4\bigr)\thetad_2
=: A_1+A_2 \thetad_2 \ .
\ea
Then plugging this into the second equation of \eqref{NL_EoM}, we have
\ba
(b_1 A_2+ b_2 + b_3 A_1)\thetad_2 = b_4 -b_1 A_1 \ .
\ea
Thus, if the uniqueness condition,
\ba
(b_1 A_2+ b_2 + b_3 A_1)^{(0)} = \bigl[a_1^{-1}(-b_1 a_2+a_1 b_2)\bigr]^{(0)}\neq 0 \ ,
\ea
is satisfied, one can solve the set of nonlinear equations \eqref{NL_EoM}. In the other case, where the equations of motion have no linear terms in $\dot{\theta}_\alpha$, there is no way to express each time derivative in terms of nonderivative variables, and the equations of motion are unsolvable. Therefore, the nonzero determinant of the coefficient matrix of the linear terms in $\thetad_\alpha$ is the necessary and sufficient condition for the equations to be uniquely solved 
for the time derivatives of the fermionic variables.

\subsubsection{More general argument}

In this subsection, we will extend the previous analysis to our degenerate theory and show
the equivalence between the condition \eqref{con:detJH}
and the condition that the fermionic equations 
\eqref{Y1} and \eqref{Y2} are uniquely solved.

The equations of motion for fermions~\eqref{Y1} and \eqref{Y2} include up to first time derivatives, and due to the Grassmann properties, they can be therefore generally written as [see also \eqref{EoM_fermion}]
\ba
\Y_\alpha &=& \sum_{n}^{2}\sum_{m}^{2} \X_{\alpha\beta_1\cdots\beta_n\gammad_1\cdots\gammad_m}
\psid^{\beta_1} \cdots \psid^{\beta_n} \psibd{}^{\gammad_1}\cdots \psibd{}^{\gammad_m}=0\,,  \label{Y1s}\\
\Y_\alphad &=& -(\Y_\alpha)^\dagger
=
-\sum_{n}^{2}\sum_{m}^{2}
\psibd{}^{\gammad_m}\cdots\psibd{}^{\gammad_1}\psid^{\beta_n} \cdots\psid^{\beta_1}   \left(\X_{\alpha\beta_1\cdots\beta_n\gammad_1\cdots\gammad_m}\right)^\dagger =0\,,\label{Y2s}
\ea
where $\X_{\alpha\beta_1\cdots\beta_n\gammad_1\cdots\gammad_m}$ consists of $\psi^\alpha$, $\psib^\alphad$, $\phi^a$, $\phid^a$, and their spatial derivatives.
If these equations can be solved for $\psid^\alpha$ and $\psibd{}^\alphad$, they should have the forms 
\begin{align}
	\dot{\psi}^\alpha&= \bigl(\dot{\bar{\psi}}^{\dot{\alpha}}\bigr)^\dagger \nonumber\\
	&=\sum_{k, l, m, n} C^{\alpha\, i_1\cdots i_m j_1\cdots j_n}_{\beta_1 \cdots \beta_k \betad_1\cdots \betad_l \delta_1 \cdots \delta_m \deltad_1 \cdots \deltad_n} \,
	\psi^{\beta_1}\cdots\psi^{\beta_k}
	\bar{\psi}^{\betad_1}\cdots\bar{\psi}^{\betad_l}
	\partial_{i_1}\psi^{\delta_1} \cdots \partial_{i_m} \psi^{\delta_m}
	\partial_{j_1}\psib^{\deltad_1} \cdots \partial_{j_n} \psib^{\deltad_n}	\ ,
 \label{ansatz}
\end{align}
where the coefficients $C^{\alpha \, i_1\cdots}_{\beta_1\cdots}$ 
consist of $\phi^a$, $\phid^a$, and $\partial_i \phi^a$ and has no dependence on $\psi^\alpha$ and $\psib^\alphad$.\footnote{$(k+l+m+n)$ is a finite odd number.}
The nonlinear equations~(\ref{Y1s}) and (\ref{Y2s}) can be solved for the first time derivatives of the fermion if we could uniquely determine the coefficients~$C^{\alpha \, i_1\cdots}_{\beta_1\cdots}$ by substituting (\ref{ansatz}) into the them.
After the substitution, 
the EOMs should become trivial at each order of ($\psi^\gamma$, $\psib^\gammad$),  ($\p_i \psi^\gamma$, $\p_i \psib^\gammad$),  ($\psi^\gamma \psi^\delta \psib^\gammad, \psi^\gamma \psib^\gammad \psib^\deltad$), and so on. At the lowest order, where the terms are linear in $\psi^\gamma$ and $\psib^\gammad$, we have
\begin{align}
	\begin{pmatrix}
		\X^{(0)}_{\alpha\beta} & \X^{(0)}_{\alpha\dot{\beta}} \\
		-\left(\X^{(0)}_{\alpha \dot{\beta}}\right)^\ast & -\left(\X^{(0)}_{\alpha\beta}\right)^\ast
	\end{pmatrix}
	\begin{pmatrix}
		C^\beta{}_{\gamma}& C^\beta{}_{\dot{\gamma}}\\
		\left(C^\beta{}_{\dot{\gamma}}\right)^\ast & \left(C^\beta{}_\gamma\right)^\ast
	\end{pmatrix}
	\begin{pmatrix}
		\psi^\gamma\\
		\bar{\psi}^{\dot{\gamma}}
	\end{pmatrix}
	=-
	\begin{pmatrix}
		\displaystyle  \left(\frac{\partial \X_{\alpha}}{\partial \psi^\gamma}\right)^{(0)} & \displaystyle\left(\frac{\partial\X_{\alpha}}{\partial \bar{\psi}^{\dot{\gamma}}}\right)^{(0)}\\
		\displaystyle -\left(\frac{\partial\X_{\alpha}}{\partial \psib^\gammad}\right)^{(0)\ast} & \displaystyle -\left(\frac{\partial \X_{\alpha}}{\partial \psi^{\gamma}}\right)^{(0)\ast}
	\end{pmatrix}
	\begin{pmatrix}
		\displaystyle  \psi^\gamma\\
		\displaystyle  \bar{\psi}^{\dot{\gamma}} 
	\end{pmatrix}\ .
	\label{Det_coeff_first}
\end{align}
Therefore, if the matrix 
\ba
{\bf J}_L^{(0)}(t,{\bf x})= -
\begin{pmatrix}
	\X^{(0)}_{\alpha\beta} & \X^{(0)}_{\alpha\dot{\beta}} \\
	-\left(\X^{(0)}_{\alpha \dot{\beta}}\right)^\ast & -\left(\X^{(0)}_{\alpha\beta}\right)^\ast  
\end{pmatrix}
\ea
has a nonzero determinant, 
we can uniquely determine $C^{\beta}{}_\gamma$ and $C^{\beta}{}_\gammad$ from \eqref{Det_coeff_first}. Similarly, 
the coefficients associated with the spatial derivative terms 
	($\p_i \psi^\gamma$, $\p_i \psib^\gammad$) are determined 
	by 
\begin{align}
\begin{pmatrix}
\X^{(0)}_{\alpha\beta} & \X^{(0)}_{\alpha\dot{\beta}} \\
-\left(\X^{(0)}_{\alpha \dot{\beta}}\right)^\ast & -\left(\X^{(0)}_{\alpha\beta}\right)^\ast
\end{pmatrix}
\begin{pmatrix}
C^{\beta i}{}_{\gamma}& C^{\beta i}{}_{\dot{\gamma}}\\
\left(C^{\beta i}{}_{\dot{\gamma}}\right)^\ast & \left(C^{\beta i}{}_\gamma\right)^\ast
\end{pmatrix}
\begin{pmatrix}
\p_i \psi^\gamma\\
\p_i \bar{\psi}^{\dot{\gamma}}
\end{pmatrix}
=-
\begin{pmatrix}
\displaystyle  \left(\frac{\partial \X_{\alpha}}{\partial (\p_i \psi^\gamma)}\right)^{(0)} & \displaystyle\left(\frac{\partial\X_{\alpha}}{\partial (\p_i  \bar{\psi}^{\dot{\gamma}})}\right)^{(0)}\\
\displaystyle -\left(\frac{\partial\X_{\alpha}}{\partial (\p_i  \psib^\gammad)}\right)^{(0)\ast} & \displaystyle -\left(\frac{\partial \X_{\alpha}}{\partial (\p_i  \psi^{\gamma})}\right)^{(0)\ast}
\end{pmatrix}
\begin{pmatrix}
\displaystyle  \p_i \psi^\gamma\\
\displaystyle  \p_i \bar{\psi}^{\dot{\gamma}} 
\end{pmatrix}\ .
\label{Det_coeff_firsts}
\end{align}
Once the coefficients at the linear order in the fermionic fields are determined, 
the coefficients at the cubic order can be determined by the equation
\begin{align}
\begin{pmatrix}
\X^{(0)}_{\alpha\beta} & \X^{(0)}_{\alpha\dot{\beta}} \\
-\left(\X^{(0)}_{\alpha \dot{\beta}}\right)^\ast & -\left(\X^{(0)}_{\alpha\beta}\right)^\ast
\end{pmatrix}
\begin{pmatrix}
&C^\beta{}_{\gamma\delta\dot{\eta}}& C^\beta{}_{\gamma\dot{\delta}\dot{\eta}} \\
& \left(C^\beta{}_{\eta\dot{\delta}\dot{\gamma}}\right)^\ast& \left(C^\beta{}_{\eta\delta\dot{\gamma}}\right)^\ast
\end{pmatrix}
\begin{pmatrix}
\psi^\gamma\psi^\delta\bar{\psi}^{\dot{\eta}} \\
\psi^\gamma\bar{\psi}^{\dot{\delta}}\bar{\psi}^{\dot{\eta}} \\
\end{pmatrix}
=-
\begin{pmatrix}
\G_{\alpha\gamma\delta{\dot \eta}} & \G_{\alpha\gamma\deltad{\dot \eta}} \\
-\left(\G_{\alpha\eta\deltad\gammad}\right)^\ast& -\left(\G_{\alpha \eta\delta\gammad} \right)^\ast
\end{pmatrix} 
\begin{pmatrix}
\psi^\gamma\psi^\delta\bar{\psi}^{\dot{\eta}} \\
\psi^\gamma\bar{\psi}^{\dot{\delta}}\bar{\psi}^{\dot{\eta}} \\
\end{pmatrix} \ .
\end{align}
Here, the coefficient matrices $\G_{\alpha\gamma\delta{\dot \eta}}$ and
$\G_{\alpha\gamma\deltad{\dot \eta}}$ are expressed in terms of
$\X_{\alpha\beta_1\cdots\beta_n\gammad_1\cdots\gammad_m}$ and
their derivatives with respect to $\psi^{\alpha}$,
$\bar{\psi}^{\dot{\alpha}}$, and their spatial derivatives, as well as
the linear coefficients $C^\beta{}_{\gamma}$, $C^{\beta}{}_{\dot{\gamma}}$, $C^{\beta i}{}_{\gamma}$, and $C^{\beta i}{}_{\dot{\gamma}}$.  Again, we can solve the
above equations for $C^\beta{}_{\gamma\delta\dot{\eta}}$ and
$C^\beta{}_{\gamma\deltad\dot{\eta}}$ and successively
determine all the coefficients appearing in \eqref{ansatz} with the
same procedure as long as the matrix
${\bf J}_L^{(0)}$ has an inverse. 

Now, it is straightforward 
to calculate the matrix ${\bf J}_L^{(0)}$ from the concrete expression of the equations of motion~\eqref{Y1} and \eqref{Y2}. The definition of ${\bf J}_L^{(0)}$ is rewritten with ${\cal Y}_\alpha$ and ${\cal Y}_\alphad$ as
\ba
{\bf J}_L^{(0)}(t,{\bf x})
=-\left(
\begin{array}{cc}
	\displaystyle	\left({\p \Y_\alpha \over \p \psid^\beta}\right)^{(0)}  & \displaystyle{\left(\p \Y_\alpha \over \p \psibd{}^\betad\right)^{(0)} }  \\
\displaystyle	{\left(\p \Y_\alphad \over \p \psid^\beta\right)^{(0)} }  & \displaystyle{\left(\p \Y_\alphad \over \p \psibd{}^\betad\right)^{(0)} } \\
\end{array}
\right) \, ,
\ea
and
the components are explicitly given by
\ba
\left({\p \Y_\alpha \over \p \psid^\beta}\right)^{(0)} &=& 
- \left( {\p \Y_\alphad \over \p \psibd{}^\betad}\right)^{\dagger(0)}
=-\Bigl(2(\SII_a)^{(0)} \p_i \phi^a-\p_i W_3^{(0)}\Bigr)(\sigma^{0i}\eps)_{\alpha\beta} \,,\\
\left({\p \Y_\alpha \over \p \psibd{}^\betad}\right)^{(0)}  &=& 
-\left( {\p \Y_\alphad \over \p \psid^\beta}\right)^{\dagger(0)}
=2(\PII)^{(0)} \sigma^0_{\alpha\betad} \,.
\ea
It is manifest that ${\bf J}_L^{(0)}$ agrees with the matrix~\eqref{JH0} with \eqref{C0},
obtained in the Hamiltonian formulation. Therefore, the condition that the fermionic equations can be
uniquely solved in the Lagrangian formulation is equivalent to the condition that 
all the primary constraints are second class. 

%%%%%%%%%%%%%%%%%%%%%%%%%%%%%%%%%%%%%
\section{Field redefinition}
\label{sec:VI}
%%%%%%%%%%%%%%%%%%%%%%%%%%%%%%%%%%%%%

So far, we have successfully obtained the most general quadratic Lagrangian for $n$-scalar and one Weyl fields satisfying the maximally degenerate conditions including up to first derivatives. 
However, one needs to carefully check whether the obtained theories can be mapped into 
a known theory, which trivially satisfies the degeneracy conditions.\footnote{Degeneracy of fermionic fields is always necessary to avoid negative norm states even if it includes only first derivatives. This fact is in sharp contrast with the cases of bosonic fields, which can be healthy as far as the Lagrangian does not include second or higher derivatives even if it is nondegenerate. When the Lagrangian does not have mixing between bosonic and fermionic fields, the maximally degenerate conditions~\eqref{MDC1} and \eqref{MDC2} are equivalent to the degeneracy of the kinetic matrix of the fermionic fields, $\det D=0$.  In a special case where each field has no mixing with the others, all the components of the submatrix~$D$ vanish; i.e., the kinetic matrix is trivially degenerate. One of the simplest examples is a non-derivative interacting system of a canonical scalar field and a Weyl field like \eqref{canonical}, which is linear in the first derivative of fermions.}
For an example, let us consider the Lagrangian containing a canonical scalar field and a Weyl field,
\ba
\Lag &=& {1\over 2} (\p_\mu \phi)^2 
+ {i\over2} \left(
\psib^\alphad \sigma_{\alpha\alphad}^\mu \p_\mu \psi^\alpha 
-\p_\mu \psib^\alphad \sigma_{\alpha\alphad}^\mu \psi^\alpha
\right) \,.
\label{canonical}
\ea
Note that the kinetic matrix of \eqref{canonical} is trivially degenerate since
${\cal B}={\cal C}=D=0$.
Under the following field redefinition, 
\ba
\phi = \varphi -{i \over 2} \Psi + {i \over 2} \Psib \,,
\label{FRD0}
\ea
where $\varphi$ is a new scalar field, while keeping the Weyl field the same,
the Lagrangian is transformed as 
\ba
\Lag &=& {1\over 2} (\p_\mu \varphi)^2 
-i \p^\mu \varphi (\psi^\alpha \p_\mu \psi_\alpha- \psib_{\alphad} \p_\mu \psib^\alphad)
+ {i\over2} \left(
\psib^\alphad \sigma_{\alpha\alphad}^\mu \p_\mu \psi^\alpha 
-\p_\mu \psib^\alphad \sigma_{\alpha\alphad}^\mu \psi^\alpha
\right)\nonumber\\
&&-{1\over 2}(\psi^\alpha \p_\mu \psi_\alpha)^2 + (\psi^\alpha \p_\mu \psi_\alpha)(\psib_\alphad \p^\mu \psib^\alphad)-{1\over 2}(\psib_\alphad \p_\mu \psib^\alphad)^2 \ .
\label{LagFT}
\ea
As one can see, the submatrices ${\cal B}, {\cal C}$, and $D$ in the kinetic matrix become nonzero due to the transformation \eqref{FRD0}, and the mapped Lagrangian apparently yields the second-order differential equations 
for the fermionic fields. One can, however, easily check that this mapped Lagrangian satisfies the maximally 
degenerate conditions~\eqref{MDC1_single} and \eqref{MDC2_single} as well as the supplementary condition~\eqref{con:detJH}; therefore, the number of d.o.f. is the same as the original Lagrangian~\eqref{canonical}.\footnote{The Hamiltonian analysis of this Lagrangian has already been examined in Ref.~\cite{Kimura:2017gcy}.} This is because the field redefinition~\eqref{FRD0} is an invertible transformation, keeping the number of d.o.f. under the transformation~\cite{Takahashi:2017zgr}.

Now, we would like to know whether our degenerate Lagrangian can be mapped into theories of which the kinetic matrix is trivially degenerate by the covariant field redefinition~$(\phi^a,\psi^\alpha,\bar{\psi}^{\dot{\alpha}})\rightarrow(f^A(\phi,\psi,\bar{\psi}), \eta^\Lambda(\phi,\psi,\bar{\psi}),\bar{\eta}^{\dot{\Lambda}}(\phi,\psi,\bar{\psi}))$,\footnote{One can consider field redefinitions including derivatives of the scalar and fermionic fields, but, for instance, redefinitions like $(\phi,\psi,\psib) \rightarrow (f(\phi,\partial \phi, \psi, \bar{\psi}),\eta(\phi, \psi, \bar{\psi}),\bar{\eta}(\phi, \psi, \bar{\psi}))$, $(\phi,\psi,\psib)\rightarrow(f(\phi,\psi,\psib),\eta(\phi,\psi,\psib,\partial\psi),\bar{\eta}(\phi,\psi,\psib,\partial\psib))$, $(\phi,\psi,\psib)\rightarrow(f(\phi,\psi,\psib),\eta(\phi,\psi,\psib,\partial\psib),\bar{\eta}(\phi,\psi,\psib,\partial\psi))$, or $(\phi,\psi,\psib)\rightarrow(f(\phi,\psi,\psib,\partial \psi,\partial\psib),\eta(\phi,\psi,\psib),\bar{\eta}(\phi,\psi,\psib))$ result in the appearance of the higher derivatives, implying that they might not keep the Lagrangian including just up to the first derivatives. To introduce such transformations, we also need to check the invertibility of them.} i.e., if it is possible that 
\ba
{\bf K}&=&
\left(
\begin{array}{ccc}
	A_{ab} & \B_{a\beta,J} & \B_{a\betad,J}\\
	\C_{\alpha b,I} & D_{\alpha\beta,IJ} & D_{\alpha\betad,IJ}\\
	\C_{\alphad b,I} & D_{\alphad\beta,IJ} & D_{\alphad\betad,IJ}\\
\end{array}
\right)
\xrightarrow{\textrm{field redefinition}}
\left(
\begin{array}{ccc}
	{\tilde A}_{ab} & 0 &0\\
	0&0 & 0\\
	0 & 0 & 0\\
\end{array}
\right), 
\label{trans:kinetic}
\ea
where ${\tilde A}_{ab} $ is the kinetic matrix of the redefined scalar fields~$f^A$.
We note that, under the field redefinition, the Lagrangians~\eqref{L0},
\eqref{L1}, and \eqref{Lag:L012}, are not mixed since the
transformation does not include the derivatives of the fields. From the
definition of the kinetic matrix, the quadratic terms in derivatives in
the Lagrangian~\eqref{Lag:L012} can be responsible for the kinetic matrix,
while linear or lower ones~\eqref{L1}, \eqref{L0} are obviously
not. Furthermore, among the quadratic terms, antisymmetrically
contracted (with respect to Lorentzian indices of the derivatives) ones
are irrelevant. The only contribution to the kinetic matrix is coming
from the symmetrically contracted ones in ${\cal L}_2$.
When all the submatrices except ${\tilde A}_{ab}$ in the kinetic matrix vanish, 
the corresponding symmetrically contracted covariant terms in the Lagrangian also should vanish simultaneously because of the covariance of the theory. 
As a result, we can just focus on the symmetrically contracted terms in ${\cal L}_2$ in this section. Picking up these, we rewrite
the relevant Lagrangian with the degeneracy conditions as 
\begin{align}
{\mathcal L}_{\rm rel}
=&\frac{1}{2}\left[V_{ac}\partial_\mu\phi^c-\frac{1}{2}\left(S_a^{(1)}\partial_\mu \Psi+(S_a^{(1)})^\dagger\partial_\mu\bar{\Psi}\right)\right]
V^{ab}
\left[V_{bd}\partial^\mu\phi^d-\frac{1}{2}\left(S_b^{(1)}\partial^\mu \Psi+(S_b^{(1)})^\dagger\partial^\mu\bar{\Psi}\right)\right]\,.
\label{Lag_rel_simple}
\end{align}
If the coefficients $V_{ab}$ and $S_a^{(1)}$ are constants, 
the square brackets $[\cdots]$ can be redefined as the derivatives of new scalar fields.
Then, the nontrivial terms in \eqref{Lag_rel_simple}, which yield second time derivatives of the fermionic fields in Euler-Lagrange equations, can be removed just as a nontrivial Lagrangian~\eqref{LagFT} is  reduced to a trivial Lagrangian~\eqref{canonical} by the transformation~\eqref{FRD0}. 
Under the field redefinition, 
$(\phi^a,\psi^\alpha,\bar{\psi}^{\dot{\alpha}})\rightarrow(f^A(\phi,\psi,\bar{\psi}), \eta^\Lambda(\phi,\psi,\bar{\psi}),\bar{\eta}^{\dot{\Lambda}}(\phi,\psi,\bar{\psi}))$,
the above Lagrangian is rewritten in terms of the new variables as
\begin{align}
 {\cal L}_{\rm rel}=&\frac{1}{2}{\cal G}^a_A V_{ab}{\cal G}^b_B\partial_\mu f^A \partial^\mu f^B
-{\cal G}^a_A V_{ab} {\cal G}^b_\Lambda \partial_\mu f^A \partial^\mu \eta^\Lambda
-{\cal G}^a_A V_{ab} \bar{{\cal G}}^b_{\dot{\Lambda}} \partial_\mu f^A \partial^\mu \bar{\eta}^{\dot{\Lambda}} \nonumber\\
&
-\frac{1}{2}{\cal G}^a_\Lambda V_{ab} {\cal G}^b_\Sigma \partial_\mu \eta^\Lambda \partial^\mu \eta^\Sigma
-{\cal G}^a_\Lambda V_{ab} \bar{{\cal G}}^b_{\dot{\Sigma}} \partial_\mu \eta^\Lambda \partial^\mu \bar{\eta}^{\dot{\Sigma}}
-\frac{1}{2}\bar{{\cal G}}^a_{\dot{\Lambda}} V_{ab} \bar{{\cal G}}^b_{\dot{\Sigma}} \partial_\mu \bar{\eta}^{\dot{\Lambda}} \partial^\mu \bar{\eta}^{\dot{\Sigma}} \ ,
\label{rel_Lag_trans}
\end{align}
where we have defined 
\begin{align}
 & {\cal G}^a_A=\phi^a_{, f^A}-\frac{1}{2}V^{ab}\left[S_b^{(1)} \Psi_{,f^A}+\bigl(S_b^{(1)}\bigr)^\dagger \bar{\Psi}_{,f^A}\right] \ , \label{GaA_def}  \\
 & {\cal G}^a_\Lambda=\phi^a_{, \eta^\Lambda}-\frac{1}{2}V^{ab}\left[S_b^{(1)} \Psi_{,\eta^\Lambda}+\bigl(S_b^{(1)}\bigr)^\dagger \bar{\Psi}_{,\eta^\Lambda}\right]  \ ,\label{GaLambda_def}
\end{align}
$\bar{{\cal G}}^a_{\dot{\Lambda}}=-({\cal G}^a_\Lambda)^\dagger$, and they satisfy $ {\cal G}^a_A=({\cal G}^a_A)^\dagger $.
Now, let us seek the field redefinition such that the quadratic terms in the first derivative of $\eta^\Lambda$ and $\bar{\eta}^{\dot{\Lambda}}$, the second line of \eqref{rel_Lag_trans}, are removed. This becomes possible if we find a transformation realizing
\begin{align}
{\cal G}^a_\Lambda
=0 \,.
\label{condition:FRD}
\end{align}
To find such a transformation, 
we first expand $V^{ab}$ and $S_a^{(1)}$ as 
\begin{align}
 V^{ab}=v^{ab (0)}+v^{ab(1)}\Psi+\bigl(v^{ab(1)}\bigr)^\ast\bar{\Psi}+v^{ab(2)}\Psi\bar{\Psi} \ ,\quad  S_a^{(1)}=s_{a}^{(0)}+s_{a}^{(1)}\bar{\Psi} \ ,
\end{align}
where $v^{ab (0)}$ and $s_{a}^{(i)}$ depend only on $\phi^a$.
Then, Eq.~\eqref{condition:FRD} can be rewritten as
\begin{align}
\phi^a_{, \eta^\Lambda}
 =&\frac{1}{2}\biggl[v^{ab(0)}s_{b}^{(0)}\Psi_{,\eta^{\Lambda}}
+v^{ab(0)}\bigl(s_{b}^{(0)}\bigr)^\ast\bar{\Psi}_{,\eta^{\Lambda}}
+\left(v^{ab(0)}s_{b}^{(1)}+\bigl(v^{ab(1)}\bigr)^\ast s_b^{(0)}\right)\Psi_{,\eta^{\Lambda}}\bar{\Psi}\nonumber\\
&\quad+\left(v^{ab(0)} \bigl(s_{b}^{(1)}\bigr)^*+v^{ab(1)}  \bigl(s_b^{(0)}\bigr)^*\right)
\Psi \Psib_{,\eta^\Lambda}
\biggr]
\label{calH_expand0}\\
 =&\frac{1}{2}\biggl[v^{ab(0)}s_{b}^{(0)}\Psi_{,\eta^{\Lambda}}
 +v^{ab(0)}\bigl(s_{b}^{(0)}\bigr)^\ast\bar{\Psi}_{,\eta^{\Lambda}}
 +\Re\left(v^{ab(0)}s_{b}^{(1)}+\bigl(v^{ab(1)}\bigr)^\ast s_b^{(0)}\right)(\Psi\bar{\Psi})_{,\eta^{\Lambda}}\nonumber\\
 &\quad+i\Im\left(v^{ab(0)}s_{b}^{(1)}+\bigl(v^{ab(1)}\bigr)^\ast s_b^{(0)}\right)(\Psi_{,\eta^\Lambda}\bar{\Psi}-\Psi\bar{\Psi}_{,\eta^\Lambda})\biggr] \ .
\label{calH_expand}
\end{align}
It is obvious that the first three terms can have their integral forms, 
but the last term of \eqref{calH_expand}, proportional to $\Psi_{,\eta^\Lambda}\bar{\Psi}-\Psi\bar{\Psi}_{,\eta^\Lambda}$, 
cannot be integrated as we will see below.
Let us explicitly see the transformation where $\phi^a$ coincides with the integral of the right-hand side of \eqref{calH_expand} except the last term.
Such a transformation is realized through
\begin{align}
 f^A=f^A\Bigl(\phi^b-\frac{1}{2}[A^b\Psi+\left(A^b\right)^*\bar{\Psi}+B^b\Psi\bar{\Psi}]\Bigr)\ , \quad 
 \eta^\Lambda=\eta^\Lambda(\phi^b, \psi^\beta, \psib^\betad) \ , \quad 
 \bar{\eta}^{\dot{\Lambda}}=\bar{\eta}^{\dot{\Lambda}}(\phi^b, \psi^\beta, \psib^\betad) \ , 
\label{trans_f_eta}
\end{align}
where $A^b$ and $B^b$ are functions of $\phi^a$, determined properly in the following.
For keeping the equivalence between the former and the latter Lagrangians, we assume the field redefinition invertible. The assumption of the invertible transformation requires the bosonic part of the whole Jacobian matrix,
\begin{align}
\textrm{Jacobian matrix}=
 \begin{pmatrix}
 \displaystyle{\frac{\partial f^A}{\partial \phi^b}} & \displaystyle{\frac{\partial f^A}{\partial \psi^\beta} }& \displaystyle{\frac{\partial f^A}{\partial \psib^\betad}}\\
 \displaystyle{\frac{\partial \eta^\Lambda}{\partial \phi^b}} & \displaystyle{\frac{\partial \eta^\Lambda}{\partial \psi^\beta}} & \displaystyle{\frac{\partial \eta^\Lambda}{\partial \psib^\betad}}\\
\displaystyle{ \frac{\partial \bar{\eta}^{\dot{\Lambda}}}{\partial \phi^b}} & \displaystyle{\frac{\partial \bar{\eta}^{\dot{\Lambda}}}{\partial \psi^\beta}} & \displaystyle{\frac{\partial \bar{\eta}^{\dot{\Lambda}}}{\partial \psib^\betad} }
 \end{pmatrix} \ ,
\end{align}
to have a nonzero determinant. The off-diagonal bosonic parts of the Jacobian matrix are inevitably zero because of the Grassmann-odd property. Therefore, the above requirement is equivalent to 
\begin{align}
 \det \Bigl(\frac{\partial f^A}{\partial \phi^b}\Bigr)^{(0)} \neq 0 \ , \qquad {\rm and} \qquad 
 \det\begin{pmatrix}
 \displaystyle{\frac{\partial \eta^\Lambda}{\partial \psi^\beta}} & \displaystyle{\frac{\partial \eta^\Lambda}{\partial \psib^\betad}}\\
 \displaystyle{\frac{\partial \bar{\eta}^{\dot{\Lambda}}}{\partial \psi^\beta}} & \displaystyle{\frac{\partial \bar{\eta}^{\dot{\Lambda}}}{\partial \psib^\betad} }
 \end{pmatrix}^{(0)}\neq 0 \ .
\end{align}
Since we have assumed the transformation is invertible, 
we can invert the definition of $f^A$ in \eqref{trans_f_eta} for $\phi^a$,
\begin{align}
 \phi^a=\frac{1}{2}\left[A^a\Psi+\left(A^a\right)^*\bar{\Psi}+B^a\Psi\bar{\Psi}\right]+\phi^{a(0)}(f) \ , \label{trans}
\end{align}
where $\phi^{a(0)}(f)$ is the inverse function of $f^A$ in \eqref{trans_f_eta}.
The derivative of \eqref{trans} with respect to $\eta^\Lambda$ gives
\begin{align}
\phi^a_{,\eta^\Lambda}
=&\frac{1}{2}\Bigl[A^a\Psi_{,\eta^\Lambda}+\left(A^a\right)^*\bar{\Psi}_{,\eta^\Lambda}+B^a\left(\Psi_{,\eta^\Lambda}\bar{\Psi}+\Psi\bar{\Psi}_{,\eta^\Lambda}\right)\Bigr]+\frac{1}{2}\phi^b_{,\eta^\Lambda}\Bigl[A^a_{,\phi^b}\Psi+\left(A^a\right)^*_{,\phi^b}\bar{\Psi}+B^a_{,\phi^b}\Psi\bar{\Psi}\Bigr] 
\nonumber\\
 =&\frac{1}{2}\biggl[A^a\Psi_{,\eta^\Lambda}+\left(A^a\right)^*\bar{\Psi}_{,\eta^\Lambda}+\Bigl(B^a+\frac{1}{2}A^b \left(A^a\right)^*_{, \phi^b}\Bigr)\Psi_{,\eta^\Lambda}\bar{\Psi}+\Bigl(B^a+\frac{1}{2}\left(A^b\right)^* A^a_{,\phi^b}\Bigr)\Psi\bar{\Psi}_{,\eta^\Lambda}\biggr] \ , 
 \label{TransD}
\end{align}
where we have recursively solved the first line and have used $\Psi_{,\eta^\Lambda}\Psi=(1/2)(\Psi\Psi)_{\eta^\Lambda}=0$ and $\bar{\Psi}_{,\eta^\Lambda}\bar{\Psi}=(1/2)(\bar{\Psi}\bar{\Psi})_{\eta^\Lambda}=0$ to find the second line without $\phi^b_{,\eta^\Lambda}$.
Now, comparing \eqref{calH_expand0} and \eqref{TransD}, we can easily determine the 
coefficients $A^a$ and $B^a$ as
\begin{align}
 A^a&=v^{ab(0)}s_b^{(0)} \ , \label{coefficientA}\\
 B^a&= \Re[C^a]\ ,  \qquad {\rm where} \qquad   C^a=v^{ab(0)}s_{b}^{(1)}+\bigl(v^{ab(1)}\bigr)^\ast s_b^{(0)}-\frac{1}{2}\left(v^{ab(0)}\bigl(s_b^{(0)}\bigr)^\ast\right)_{,\phi^c}v^{cd(0)}s_d^{(0)} \ .
\label{coefficientB}
\end{align}
Introducing the transformation to the definitions~\eqref{GaA_def} and \eqref{GaLambda_def}, we then have
\begin{align}
 {\cal G}^a_\Lambda
&=-\frac{i}{2} \Im[C^a]\bigl(\Psi_{,\eta^\Lambda}\bar{\Psi}-\Psi\bar{\Psi}_{,\eta^\Lambda}\bigr) \ ,
\label{GaLambda} \\
{\cal G}^a_A &=
\Bigl(\delta^a_b+\frac{1}{2}\bigl(A^a_{,\phi^b}\Psi+(A^\ast)^a_{,\phi^b}\bar{\Psi}\bigr)+\frac{1}{2}\bigl(B^a_{,\phi^b}+\Re [A^a_{,\phi^c}(A^\ast)^c_{,\phi^b}]\bigr)\Psi\bar{\Psi}\Bigr)\phi^{b(0)}_{,f^A} - {i \over 2} \Im[C^a] \left(
\Psi_{,f^A}\Psib -\Psi \Psib_{,f^A}
\right) \,.
\end{align}
Since we cannot pick up the imaginary part of $C^a$ as one can see from \eqref{coefficientB},
we still have nonvanishing ${\cal G}^a_\Lambda$. 
We cannot remove the dependence on $\Psi_{,\eta^\Lambda}\bar{\Psi}-\Psi\bar{\Psi}_{,\eta^\Lambda}$ in any way
as we expected from the expression of \eqref{calH_expand}.  
Nevertheless, the quadratic derivative interactions of fermions in \eqref{rel_Lag_trans}
is the square of \eqref{GaLambda} (and its Hermitian), and thus they accidentally vanish due to the Grassmann properties,
\begin{align}
	\mathcal{G}^a_{\Lambda}\mathcal{G}^b_{\Sigma} \propto (\Psi_{,\eta^\Lambda}\bar{\Psi}-\Psi\bar{\Psi}_{,\eta^\Lambda})(\Psi_{,\eta^\Sigma}\bar{\Psi}-\Psi\bar{\Psi}_{,\eta^\Sigma})&=0 \ , \\
	\mathcal{G}^a_{\Lambda}\bar{\mathcal{G}}^b_{\dot{\Sigma}} \propto (\Psi_{,\eta^\Lambda}\bar{\Psi}-\Psi\bar{\Psi}_{,\eta^\Lambda})(\Psi_{,\bar{\eta}^{\dot{\Sigma}}}\bar{\Psi}-\Psi\bar{\Psi}_{,\bar{\eta}^{\dot{\Sigma}}})&=0 \ ;
\end{align}
whereas, the cross term of the redefined scalar field and fermionic field cannot be 
removed because
\ba
{\cal G}^a_A V_{ab} {\cal G}^b_\Lambda \partial_\mu f^A \partial^\mu \eta^\Lambda
=- {i \over 2}v_{ab}^{(0)} \phi^{a(0)}_{,f^A} \Im[C^b] \left(
\Psi_{,\eta^\Lambda}\Psib -\Psi \Psib_{,\eta^\Lambda}
\right)
 \partial_\mu f^A \partial^\mu \eta^\Lambda\neq 0\,.
\ea
As a result, the final Lagrangian with the field redefinition \eqref{trans_f_eta} with \eqref{coefficientA} and \eqref{coefficientB} is given by
\begin{align}
 {\cal L}_{\rm rel}=&\frac{1}{2}{\cal G}^a_A V_{ab}{\cal G}^b_B\partial_\mu f^A \partial^\mu f^B-{\cal G}^a_A V_{ab} {\cal G}^b_\Lambda \partial_\mu f^A \partial^\mu \eta^\Lambda
-{\cal G}^a_A V_{ab} \bar{{\cal G}}^b_{\dot{\Lambda}} \partial_\mu f^A \partial^\mu \bar{\eta}^{\dot{\Lambda}} \, .
\end{align}
Therefore, we conclude that the fermionic derivative interactions like $\partial_\mu \eta^\Lambda \partial^\mu \eta^\Sigma$ can be eliminated by the field redefinition, but the cross terms between scalar fields and fermions like $\partial_\mu f^A \partial^\mu \eta^\Lambda$ cannot be removed.
Such cross terms indicate that the Euler-Lagrange equations apparently contain the second derivatives of the fermionic field, while they are reduced to first-order equations because the maximally degenerate conditions are satisfied,
\begin{align}
 &D_{\Lambda\Sigma}-\C_{\Lambda B}A^{BA}\B_{A\Sigma}=\G^a_B V_{ab}\G^b_{\Lambda}A^{BA} \G^c_A V_{cd}\G^d_{\Sigma}\propto \G^b_{\Lambda}\G^d_{\Sigma}=0\ , \\
 &D_{\Lambda\dot{\Sigma}}-\C_{\Lambda B}A^{BA}\B_{A\dot{\Sigma}}=
 \G^a_B V_{ab}\G^b_{\Lambda}A^{BA} \G^c_A V_{cd} \bar{\G}^d_{\dot{\Sigma}}\propto \G^b_{\Lambda}\bar{\cal G}^d_{\dot{\Sigma}}=0\ .
\end{align}
To summarize, the kinetic matrix cannot be transformed to the form in \eqref{trans:kinetic} with any field redefinition but it can be reduced to
\ba
{\bf K}&=&
\left(
\begin{array}{ccc}
	A_{ab} & \B_{a\beta,J} & \B_{a\betad,J}\\
	\C_{\alpha b,I} & D_{\alpha\beta,IJ} & D_{\alpha\betad,IJ}\\
	\C_{\alphad b,I} & D_{\alphad\beta,IJ} & D_{\alphad\betad,IJ}\\
\end{array}
\right)
\xrightarrow{\textrm{field redefinition}}
\left(
\begin{array}{ccc}
	{\tilde A}_{ab} & \tilde{\B}_{a\beta,J} & \tilde{\B}_{a\betad,J}\\
	\tilde{\C}_{\alpha b,I} & 0 & 0\\
	\tilde{\C}_{\alphad b,I} & 0 & 0
\end{array}
\right).
\ea
This fact indicates that the cross terms are really newly found derivative interactions.

The crucial difference between the purely bosonic degenerate system (such as
Horndeski theory
\cite{Horndeski:1974wa,Deffayet:2011gz,Kobayashi:2011nu} and Degenerate Higher-Order Scalar-Tensor (DHOST)
theories \cite{Langlois:2015cwa,BenAchour:2016fzp}) and our
scalar-fermion degenerate system is the highest derivatives appearing
in the Lagrangian. This is because the Lagrangian containing first
derivatives of the fermionic fields in general yields second-order
differential equations of motion, which is a signal of fermionic
(Ostrogradsky's) ghosts.  Therefore, this situation corresponds to the
bosonic Lagrangian containing up to second derivatives.  On the analogy of the conformal and disformal transformation in the scalar-tensor theory, the fermionic version of conformal (disformal)
transformations does not involve any derivatives of the fermionic
field.
Thus, the conformal transformation, which mixes the metric and the Weyl field,
takes the following form\footnote{
A candidate for the disformal transformation including the Weyl field is 
\ba
{\bar g}_{\mu\nu} = A(\Psi, \Psib) g_{\mu\nu} +B(\Psi, \Psib) J_\mu J_\nu\,, \nonumber
\ea
where $J_\mu$ is now appropriately mapped with tetrad fields $e_\mu^{~a}$, defined by $g_{\mu\nu}=e_\mu^{~a} e_\nu^{~b} \eta_{ab}$, from a flat tangent space.
However, one can easily show that the arbitrary function $B$ can be absorbed into the conformal factor $A$ by using \eqref{JJ}. 
On the other hand, the identity~\eqref{id_multi1} tells us that 
the disformal transformation with multiple Weyl fields is independent from the conformal transformation. 
}:
\ba
{\bar g}_{\mu\nu} = A(\Psi, \Psib) g_{\mu\nu} \,.
\ea
Such a transformation will be significantly useful in finding new theories of
``tensor-fermion theories,'' and we will leave this interesting issue to future work. 

%%%%%%%%%%%%%%%%%%%%%%%%%%%%%%%%%%%%%
\section{Summary and discussion}
\label{sec:summary}
%%%%%%%%%%%%%%%%%%%%%%%%%%%%%%%%%%%%%

As usually discussed in the Lagrangian composed of bosonic d.o.f., we have to avoid the appearance of ghosts even in boson-fermion
coexisting Lagrangians. Fermions easily suffer from the fermionic ghost
as pointed out in Ref.~\cite{Henneaux:1992ig}; i.e., fermionic d.o.f. should be constrained by the same number of constraints as the
(physical) d.o.f. That is true even when we additionally have bosonic
d.o.f.

Our purpose is the construction of covariant derivative interactions
between fermions or between scalar fields and fermions, free from
fermionic ghosts.  We have given the prescription to find new
interactions in the Lagrangian with up to the first derivative of fields, made
up of a set of conditions imposed on the Lagrangian. One of the conditions
is the maximally degenerate condition which produces the sufficient number of
primary constraints. In this paper, all of the constraints are
simply assumed to be second class. If we have gauge invariance, that is,
first class constraints, the number of primary constraints can be
smaller. However, it should be notice that, after the gauge fixing, all of
the primary constraints can be turned into second class. In the Lagrangian
formulation, the equations of motion for fermions can be reduced to
the first-order differential equations by imposing the maximally
degenerate condition, which should be solved for the first
derivative of each fermionic variable. Whether they are solvable is not
{\it a priori} because of the Grassmann-odd nature, and the requirement is
exactly the same with the invertibility of the constraint matrix.

The coefficients of the derivative of fields in the Lagrangian are
complicated but classified thanks to the covariance, and they become
much simpler in case of one Weyl field with multiple scalar fields.
One of the prominent features which we can learn from the concrete
analysis is that the first condition, that is, the maximally degenerate
condition, affects the symmetric part of the Lagrangian with respect to
the space-time indices of the derivatives, but the supplementary second
condition affects the antisymmetric part.

The proposed Lagrangian
is possible to be used as a model for the interaction between scalar
fields and a Weyl field, but some may ask what happens when we perform
field redefinitions. Transformation including derivatives seems to
introduce cubic or higher nonlinear derivative terms as well as second
or higher derivatives in the Lagrangian. In a simple setup where the
whole Lagrangian is expressed by the proposed quadratic Lagrangian, the
quadratic terms in the derivative of fermions are absorbed by certain
invertible field redefinitions, but the derivative interaction terms
between the derivative of the scalar fields and the fermions remain,
which suggests that they are strictly new terms.
As shown in Ref.~\cite{Motohashi:2017eya}, in a bosonic case, any healthy
theory even with arbitrary higher derivative terms can be reduced to a
nondegenerate Lagrangian written in terms of (unconstrained) variables
with at most the first-order time derivatives through canonical
transformation. On the other hand, in a fermionic case, any healthy
theories require constraints to reduce extra d.o.f. of
fermionic variables. 
Then, it would be interesting to investigate what kind of a simpler Lagrangian can be generally found through canonical transformation from healthy fermionic theories even with arbitrary higher derivative terms.

There are several directions for applying the method we have developed:
inclusion of nonlinear derivative interactions, second or higher
derivatives, gauge invariance and/or gravity, and so on. Some of these
issues will be discussed in future publications.

%--- Acknowledgments ---%--- Acknowledgments ---%--- Acknowledgments ---%
\acknowledgments 
We would like to thank Teruaki Suyama for fruitful discussions. This work was supported in part by JSPS KAKENHI Grants No. JP17K14276 (R.K.), No. 
JP15H05888 (M.Y.), No. JP25287054 (M.Y.), and No. JP18H04579 (M.Y.).
%--- Acknowledgements ---%--- Acknowledgements ---%--- Acknowledgements ---%

\appendix

%%%%%%%%%%%%%%%%%%%%%%%%%%%%%%%%%%%%%
\section{Identities of Pauli matrices}
\label{app:A}
%%%%%%%%%%%%%%%%%%%%%%%%%%%%%%%%%%%%%

In this Appendix, we summarize notations and identities of Pauli matrices~\cite{Wess:1992cp}. 
We have used the following matrices defined as
\ba
\sigmab^{\mu \alphad\alpha}&=& \eps^{\alphad\betad}\eps^{\alpha\beta}\sigma_{\beta\betad}^\mu \,,\label{B1}\\
(\sigma^{\mu\nu})_\alpha^{~\beta}= 
{1\over 4} \(\sigma_{\alpha\alphad}^\mu \sigmab^{\nu\alphad\beta}
-\sigma_{\alpha\alphad}^\nu \sigmab^{\mu\alphad\beta} \)\ , \label{B5} &\quad&
(\sigmab^{\mu\nu})^\alphad_{~\betad}=
{1\over 4} \left(\sigmab^{\mu\alphad\alpha} \sigma^{\nu}_{\alpha\betad}
-\sigmab^{\nu\alphad\alpha} \sigma^{\mu}_{\alpha\betad} \right)\ ,\\
(\sigma^{\mu\nu} \eps)_{\alpha\beta} = (\sigma^{\mu\nu})_\alpha^{~\gamma} \, \eps_{\gamma\beta}\,,\label{B9} &\quad&
(\eps \sigmab^{\mu\nu} )_{\alphad\betad} = \eps_{\alphad\gammad} \, (\sigmab^{\mu\nu})^\gammad_{~\betad} \ .
\ea
Note that the last two matrices are symmetric under the exchange of the fermionic indices, i.e., $(\sigma^{\mu\nu} \eps)_{\alpha\beta} =(\sigma^{\mu\nu} \eps)_{\beta\alpha} $ and $(\eps \sigmab^{\mu\nu} )_{\alphad\betad}=(\eps \sigmab^{\mu\nu} )_{\betad\alphad}$.
They satisfy the following useful properties:
\ba
&&\sigma_{\alpha\alphad}^\mu \sigmab_\mu^{\betad\beta} = -2 \, \delta_\alpha^{~\beta} \delta_\alphad^{~\betad}\ , \quad
\sigma_{\alpha\alphad}^\mu \sigma_{\mu\beta\betad} = -2 \, \eps_{\alphad\betad} \, \eps_{\alpha\beta}\ , \quad \label{B3}
\sigmab^{\mu\alphad\alpha} \sigmab_\mu^{\betad\beta} =  -2 \, \eps^{\alphad\betad} \, \eps^{\alpha\beta}\ ,\\
&&\sigma_{\alpha\alphad}^\mu \sigma_{\beta\betad}^\nu-\sigma_{\alpha\alphad}^\nu \sigma_{\beta\betad}^\mu
=2\bigr[ (\sigma^{\mu\nu} \eps)_{\alpha\beta} \, \eps_{\alphad\betad}
+ (\eps \sigmab^{\mu\nu} )_{\alphad\betad} \, \eps_{\alpha\beta}  \bigr]\,,\label{B7}\\
&&\sigma_{\alpha\alphad}^\mu \sigma_{\beta\betad}^\nu+\sigma_{\alpha\alphad}^\nu \sigma_{\beta\betad}^\mu
=-\eta^{\mu\nu} \, \eps_{\alpha\beta} \, \eps_{\alphad\betad}
-4 (\sigma^{\mu\rho} \eps)_{\alpha\beta}  (\eps \sigmab_\rho{}^{\nu} )_{\alphad\betad}\ .\label{B8}
\ea
In addition, one can easily derive the following useful identities:
\begin{align}
  &(\sigma^{\mu\nu} \eps)_{\alpha\beta} \sigma_{\nu \gamma \gammad} =
  {1 \over 2} (\eps_{\alpha\gamma} \sigma^\mu_{\beta\gammad} + \eps_{\beta\gamma}\sigma^\mu_{\alpha\gammad}) \ ,\label{A7}\\
 &(\sigma^{\mu\rho}\varepsilon)_{\alpha\beta}(\varepsilon\bar{\sigma}_\rho{}^{\nu})_{\gammad\deltad}=-\frac{1}{8}\bigl(\sigma^\mu_{\alpha\gammad}\sigma^\nu_{\beta\deltad}+\sigma^\mu_{\alpha\deltad}\sigma^\nu_{\beta\gammad}+\sigma^\nu_{\alpha\gammad}\sigma^\mu_{\beta\deltad}+\sigma^\nu_{\alpha\deltad}\sigma^\mu_{\beta\gammad}\bigr) \ , \label{B10} \\
  &(\sigma^{\mu\rho} \eps)_{\alpha\beta} (\sigma_\rho{}^{\nu} \eps)_{\gamma\delta}
  = {1\over 4} \Bigl[
  -\eta^{\mu\nu} (
  \eps_{\alpha\gamma} \eps_{\beta\delta}
  +\eps_{\beta\gamma} \eps_{\alpha\delta})
  +(\sigma^{\mu\nu} \eps)_{\alpha\gamma}\eps_{\beta\delta}
  +(\sigma^{\mu\nu} \eps)_{\beta\gamma}\eps_{\alpha\delta}
  +(\sigma^{\mu\nu} \eps)_{\alpha\delta}\eps_{\beta\gamma}
  +(\sigma^{\mu\nu} \eps)_{\beta\delta}\eps_{\alpha\gamma}
   \Bigr]\ . ~~\label{A11}
\end{align}
The contraction of the Weyl indices is given as
\begin{align}
 &\eps^{\alpha\beta}\sigma^\mu_{\alpha\alphad} \sigma^\nu_{\beta\betad} 
 = -2 (\eps\sigmab^{\mu\nu})_{\alphad\betad} +\eta^{\mu\nu}\eps_{\alphad\betad} \ , \\
 &\eps^{\alphad\betad}\sigma^\mu_{\alpha\alphad} \sigma^\nu_{\beta\betad} 
 = -2 (\sigma^{\mu\nu} \eps)_{\alpha\beta} +\eta^{\mu\nu}\eps_{\alpha\beta} \ .
\end{align}
Therefore, any contraction of the building block matrices, $\eps_{\alpha\beta}$, $\eps_{\alphad\betad}$, $ \sigma_{\alpha\alphad}^\mu$, $(\sigma^{\mu\nu}\varepsilon)_{\alpha\beta}$, and $(\varepsilon\bar{\sigma}^{\mu\nu})_{\alphad\betad}$, with respect to
the Lorentzian or fermionic indices reduces to the uncontracted combination of them.
Here, Eq.~(\ref{B10}) implies
 $(\sigma^{\mu\rho}\varepsilon)_{\alpha\beta}(\varepsilon\bar{\sigma}_\rho{}^{\nu})_{\gammad\deltad}$ is symmetric under the exchange of $\mu$ and $\nu$ and traceless,
\begin{align}
 (\sigma^{\mu\rho}\varepsilon)_{\alpha\beta}(\varepsilon\bar{\sigma}{}_{\rho\mu})_{\dot{\gamma}\dot{\delta}}=0 \ ,
\end{align}
through \eqref{B3}.
The matrices and their complex conjugates are related through
\begin{align}
 &\bigl(\sigma^\mu_{\alpha\betad}\bigr)^\ast=\sigma^\mu_{\beta\alphad}\ , \quad 
 \bigl((\sigma^{\mu\nu}\varepsilon)_{\alpha\beta}\bigr)^\ast=(\varepsilon\bar{\sigma}^{\mu\nu})_{\betad\alphad} \ .
\end{align}

%%%%%%%%%%%%%%%%%%%%%%%%%%%%%%%%%%%%%
\section{Maximally degenerate conditions and primary constraints}
\label{app:B}
%%%%%%%%%%%%%%%%%%%%%%%%%%%%%%%%%%%%%
As discussed in Sec.~\ref{sec:II}, 
we need $4N$ constraints making the number of the d.o.f. of fermions half to avoid negative norm states in a $n$-scalar and $N$-fermion Lagrangian with up to first derivatives. For the purpose, we have adopted
 the maximally degenerate conditions, Eqs.~(\ref{MDC1}) and (\ref{MDC2}), having $4N$ primary constraints for the fermions, Eqs.~(\ref{PRI1}) and (\ref{PRI2}). Here, we show that, if we have $4N$ primary constraints for the fermions, the maximally degenerate conditions are automatically satisfied.

From the definition of the canonical momenta of fermions, we generally have
\ba
\pi_{\psi_I^\alpha}  = F_{\alpha,I} (\phi^a, \pi_{\phi^a} , \p_i \phi^a, \psi_J^\beta, \p_i \psi_J^\beta, \psib_J^\betad, \p_i \psib_J^\betad, \psid^\beta_J, \psibd{}^\betad_J) \,,
\label{mome1}
\ea
after we locally solve the canonical momenta of the scalar fields for $\phid^a$, which can be justified by our assumption that $\partial \pi_a/\partial \phi^b=A_{ab}$ has an inverse.
Taking the derivative of (\ref{mome1}) with respect to $\psid_J^\beta$ and $\psibd{}_J^\betad$, we have
\begin{align}
\Lag_{\psid_I^\alpha \psid_J^\beta} = \Lag_{\phid^a \psid{}_J^\beta} {\p F_{\alpha,I} \over \p \pi_{\phi^a} }+\frac{\partial F_{\alpha, I}}{\partial \psid^\beta_J}
\quad &\Leftrightarrow\quad
\frac{\partial F_{\alpha, I}}{\partial \psid^\beta_J}=D_{\alpha\beta,IJ}+\B_{a\beta,J}{\p F_{\alpha,I} \over \p \pi_{\phi^a} } 
=D_{\alpha\beta,IJ}-\C_{\alpha b,I}A^{ba}\B_{a\beta,J}\ ,
\label{Ld:eq2}\\
\Lag_{\psid_I^\alpha \psibd{}_J^\betad} = \Lag_{\phid^a \psibd{}_J^\betad} {\p F_{\alpha,I} \over \p \pi_{\phi^a} }+\frac{\partial F_{\alpha, I}}{\partial \psibd{}^\betad_J}
\quad &\Leftrightarrow\quad
\frac{\partial F_{\alpha, I}}{\partial \psibd{}^\betad_J}=D_{\alpha\betad,IJ}+\B_{a\betad,J}{\p F_{\alpha,I} \over \p \pi_{\phi^a} } 
=D_{\alpha\betad,IJ}-\C_{\alpha b,I}A^{ba}\B_{a\betad,J}\ ,
\label{Ld:eq3} 
\end{align}
where we have used the relations derived from the derivative of \eqref{mome1} with respect to $\phid^a$,
\begin{align}
 \Lag_{\psid_I^\alpha \phid^a} = \Lag_{\phid^b \phid^a} {\p F_{\alpha,I} \over \p \pi_{\phi^b} } 
\quad\Leftrightarrow\quad
\C_{\alpha a, I} = A_{b a}{\p F_{\alpha,I} \over \p \pi_{\phi^b} } \ ,
\label{Ld:eq1}
\end{align}
in the last equalities. 
Thus, Eqs.~(\ref{Ld:eq2}) and (\ref{Ld:eq3}) tell us that no dependence of $F_{\alpha, I}$ on the time derivative of the fermions means that we have $4N$ primary constraints for the fermions.
Thus, Eqs.~(\ref{Ld:eq2}) and (\ref{Ld:eq3}) suggest that the maximally degeneracy conditions, Eqs.~(\ref{MDC1}) and (\ref{MDC2}), are not only the sufficient condition for the existence of $4N$ primary constraints for fermions but also the necessary condition for that.

\section{Scalar-fermion theories quadratic in first derivatives of multiple Weyl fields}
\label{app:C}
In this Appendix, we extend our analysis in Sec.~\ref{sec:III} to the case of multiple Weyl fields and construct the most general scalar-fermion theory of which the Lagrangian 
contains up to quadratic in first derivatives of scalar and fermionic fields.
As in Sec.~\ref{sec:III}, we first construct Lorentz-invariant scalar quantities with no derivatives, which only contain scalar fields $\phi^a$ and fermionic fields $\psi^\alpha_I$, $\psib^\alphad_I$. Since the fermionic indices can be contracted with $\eps_{\alpha\beta}$, $\eps_{\alphad\betad}$, $ \sigma_{\alpha\alphad}^\mu$, $(\sigma^{\mu\nu}\varepsilon)_{\alpha\beta}$, and $(\varepsilon\bar{\sigma}^{\mu\nu})_{\alphad\betad}$, we have five possibilities,
\ba
&& 
\Psi_{IJ} = \psi_I^\alpha \psi_{J,\alpha} , \qquad  
{\bar \Psi}_{IJ} = \psib_{I,\alphad} \psib_J^\alphad, \qquad
J_{IJ}^\mu =\psib_I^\alphad \sigma_{\alpha\alphad}^\mu \psi_J^\alpha, \nonumber\\
&&K_{IJ}^{\mu\nu} = \psi^\alpha_I (\sigma^{\mu\nu}\varepsilon)_{\alpha\beta} \psi^\beta_J, \qquad
{\bar K}_{IJ}^{\mu\nu} = \psib^\alphad_I (\varepsilon\bar{\sigma}^{\mu\nu})_{\alphad\betad} \psib^\betad_J, 
\ea
which satisfy the following properties, 
\begin{align}
	&\Psi_{IJ}=\Psi_{JI}, \qquad K^{\mu\nu}_{IJ}=-K^{\mu\nu}_{JI}, \qquad K^{\mu\nu}_{IJ}=-K^{\nu\mu}_{IJ}\,,\nonumber\\
	& (\Psi_{IJ})^\dagger={\bar \Psi}_{JI}, \qquad(J_{IJ}^\mu)^\dagger = J_{JI}^\mu, \qquad (K_{IJ}^{\mu\nu} )^\dagger = {\bar K}_{JI}^{\mu\nu}\,,
\end{align}
and 
\begin{align}
	&J^{\mu}_{IJ}J^{\nu}_{KL}=-\frac{1}{2}\eta^{\mu\nu}\bar{\Psi}_{IK}\Psi_{JL}
	-2\bar{K}^{\lambda\nu}_{IK}K^{\lambda\mu}_{JL}+\bar{\Psi}_{IK}K^{\mu\nu}_{JL}-\Psi_{JL}\bar{K}^{\mu\nu}_{IK}\ ,\label{id_multi1} \\
	&J^{\mu}_{IJ}J_{\mu \, KL}=\eta_{\mu\nu}J_{IJ}^\mu J_{KL}^\nu = -2{\bar \Psi}_{IK}\Psi_{JL} \ , \label{id_multi2}\\
	&K^{\mu\nu}_{IJ}J_{\nu \, KL}=\frac{1}{2}(J^{\mu}_{KI}\Psi_{JL}-J^{\mu}_{KJ}\Psi_{IL}) \ , \label{id_multi3}\\
	&K^{\mu\nu}_{IJ}J_{\nu \, KL}J_{\mu \, MN}=-\bar{\Psi}_{KM}(\Psi_{IN}\Psi_{JL}-\Psi_{JN}\Psi_{IL}) \ , \label{id_multi4}\\
	&\bar{K}^{\mu\nu}_{IJ}J_{\nu \, KL}=\frac{1}{2}(-J^{\mu}_{IL}\bar{\Psi}_{JK}+J^{\mu}_{JL}\bar{\Psi}_{IK}) \ , \label{id_multi5}\\
	&\bar{K}^{\mu\nu}_{IJ}J_{\nu \, KL}J_{\mu \, MN}=-\Psi_{NL}(\bar{\Psi}_{MJ}\bar{\Psi}_{KI}-\bar{\Psi}_{MI}\bar{\Psi}_{KJ}) \ , \label{id_multi6}\\
	&K^{\mu\nu}_{IJ}K_{\nu}{}^{\lambda}{}_{KL}=\frac{1}{4}\eta^{\mu\lambda}(-\Psi_{IL}\Psi_{JK}+\Psi_{IK}\Psi_{JL})+\frac{1}{4}(K^{\mu\lambda}_{IL}\Psi_{JK}-K^{\mu\lambda}_{IK}\Psi_{JL}-K^{\mu\lambda}_{JL}\Psi_{IK}+K^{\mu\lambda}_{JK}\Psi_{IL}) \ ,\label{id_multi7}\\
	&K^{\mu\nu}_{IJ}\bar{K}_{\nu}{}^{\lambda}{}_{KL}=\frac{1}{8}(J^{\mu}_{KI}J^\lambda_{LJ}-J^\mu_{LI}J^{\lambda}_{KJ}-J^\mu_{KJ}J^\lambda_{LI}+J^\mu_{LJ}J^\lambda_{KI})=K^{\lambda\nu}_{IJ}\bar{K}_{\nu}{}^{\mu}{}_{KL}\ , \label{id_multi8}\\
	&K^{\mu\nu}_{IJ}\bar{K}_{\nu\mu}{}_{KL}=\frac{1}{4}(J^{\mu}_{KI}J_{\mu \, LJ}-J^\mu_{LI}J_{\mu \, KJ})=0 \label{id_multi9}\ , 
\end{align}
where we have used \eqref{B3}--\eqref{A11}.
In general, any quantities without derivatives and Weyl indices (Weyl indices are appropriately contracted) are written as 
\begin{align}
	{\cal A}^{\mu_1 \mu_2 \cdots \mu_n} = {\cal A}^{\mu_1 \mu_2 \cdots \mu_n}(\eta_{\mu\nu}, \phi^a,  \Psi_{IJ},{\bar \Psi}_{IJ},J_{IJ}^\mu , K_{IJ}^{\mu\nu} ,{\bar K}_{IJ}^{\mu\nu}) \ ,
\end{align}
where $n$ is zero or a natural number.
In addition, from the above identities, we know that all scalar quantities without derivatives can be expressed only by $\phi^a$, $\Psi_{IJ}$, and $\bar{\Psi}_{IJ}$. We also see that the space-time index of vector quantities is described by $J^\mu_{IJ}$, and those of second-rank tensors are described by $\eta^{\mu\nu}$, $K^{\mu\nu}_{IJ}$, $\bar{K}^{\mu\nu}_{IJ}$ and $K^{\mu\lambda}_{IJ}\bar{K}_{\lambda}{}^{\nu}{}_{KL}$, i.e.,
\begin{align}
	{\cal F}={\cal F}(\phi^a, \Psi_{IJ},\bar{\Psi}_{IJ}) \ , \quad 
	{\cal G}^\mu={\cal G}^{IJ}_{(1)}J^\mu_{IJ} \ , \quad 
	{\cal H}^{\mu\nu}={\cal H}_{(1)}\eta^{\mu\nu}+{\cal H}_{(2)}^{IJ}K^{\mu\nu}_{IJ}+{\cal H}_{(3)}^{IJ}\bar{K}^{\mu\nu}_{IJ}+{\cal H}_{(4)}^{IJKL}K^{\mu\lambda}_{IJ}\bar{K}_\lambda{}^\nu{}_{KL}\ ,
\end{align}
where ${\cal F}$, ${\cal G}^{\mu}$, and ${\cal H}^{\mu\nu}$ are an arbitrary scalar, vector, and second-rank tensor, respectively. ${\cal G}_{(1)}$, ${\cal H}_{(1)}$, ${\cal H}_{(2)}$, and so on are scalar quantities just like ${\cal F}$. 
Though further investigation is needed for the construction of arbitrary functions with Weyl indices, we can formally write down the most general action with arbitrary functions, and this is given by 
\ba
S=\int d^4 x \, \left( {\cal L}_0 + {\cal L}_1 +{\cal L}_2 \right)\,
\label{Gene_Lag}
\ea
where 
\ba
{\cal L}_0 &=& P_0 \,, \nonumber\\
{\cal L}_1 &=& 
\PI{}^\mu_a\, \p_\mu \phi^a 
+ \PII{}^{\mu,I}_\alpha  \p_\mu\psi_I^\alpha
+  \p_\mu\psib_I^\alphad \left(\PII{}^{\mu,I}_\alpha \right)^\dagger \nonumber
\\
{\cal L}_2 &=& 
{1 \over 2} V_{ab}^{\mu\nu}\p_\mu \phi^a \,\p_\nu \phi^b
+S_{a\alpha}^{\mu\nu, I} \p_\mu \phi^a \p_\nu \psi_I^\alpha 
+ \p_\mu \phi^a \p_\nu \psib_I^\alphad  \left(S_{a\alpha}^{\mu\nu,I}\right)^\dagger \,,\nonumber\\
&&
+{1 \over 2}W^{\mu\nu,IJ}_{\alpha\beta} \,\p_\mu \psi_I^\alpha \p_\nu \psi_J^\beta 
+{1 \over 2}\,\p_\nu \psib_J^\betad \p_\mu  \psib_I^\alphad  \left(W^{\mu\nu,IJ}_{\alpha\beta} \right)^\dagger
+\p_\mu \psib_I^\alphad \, Q^{\mu\nu,IJ}_{\alpha\alphad} \, \p_\nu \psi_J^\alpha
\,, \label{Lag_multi}
\ea
and
 $P_0$, $\PI{}^\mu_a$, $\PII{}^{\mu,I}_\alpha$, $V_{ab}^{\mu\nu}$, $S_{a\alpha}^{\mu\nu, I} $, $W^{\mu\nu,IJ}_{\alpha\beta}$, and $Q^{\mu\nu,IJ}_{\alpha\alphad}$ are appropriately contracted arbitrary functions of 
Lorentz scalar quantities, Pauli matrices, and fermionic fields. 
The properties of these coefficients are 
\begin{align}
	&(P_0)^\dagger=P_0 \ , \qquad 
	(P^{(1)}{}^\mu_a)^\dagger=P^{(1)}{}^\mu_a \ , \qquad
	(V^{\mu\nu}_{ab})^\dagger=V^{\mu\nu}_{ab} \ , \qquad
	(Q^{\mu\nu,IJ}_{\alpha\dot{\beta}})^\dagger=Q^{\nu\mu,JI}_{\beta\dot{\alpha}}\ , \nonumber\\
	&V^{\mu\nu}_{ab}=V^{\nu\mu}_{ba} \ , \qquad
	W^{\mu\nu, IJ}_{\alpha\beta}=-W^{\nu\mu,JI}_{\beta\alpha} \ .
	\label{Prop_coef}
\end{align}
The maximally degenerate conditions, Eqs.~\eqref{MDC1} and \eqref{MDC2}, are
\begin{align} 
	&D_{\alpha\beta,IJ}-\C_{\alpha b,I} A^{ba} \B_{a\beta,J}
	=W_{\alpha\beta}^{00, IJ}+S_{b\alpha}^{00,I}(V^{00})^{-1 \, ba}S_{a\beta}^{00, J}
	=0 \ , \\
	&D_{\alpha\betad,IJ}-\C_{\alpha b,I} A^{ba} \B_{a\betad,J}
	=-Q^{00,JI}_{\alpha\dot{\beta}}-S^{00,I}_{b\alpha}(V^{00})^{-1 \, ba}(S_{a\beta}^{00,J})^\dagger =0 \ .\label{Max_gene}
\end{align}
We can write down the momenta as 
\begin{align}
	\pi_{\phi^a}&=P^{(1)0}_a+V^{0\nu}_{ab}\partial_\nu\phi^b+S^{0\nu, I}_{a\alpha}\partial_\nu\psi^\alpha_I+\partial_\nu\bar{\psi}^\alphad_I(S^{0\nu, I}_{a\alpha})^\dagger \ ,\label{Mom_phi_multi}\\
	\pi_{\psi^\alpha}^I&=-P^{(2)0,I}_\alpha-S^{\mu 0,I}_{a\alpha}\partial_\mu\phi^a+W^{0\nu,IJ}_{\alpha\beta}\partial_\nu\psi^\beta_J-\partial_\mu\bar{\psi}^\betad
	Q^{\mu0,JI}_{\alpha\betad} \ ,\label{Mom_psi_multi}\\
	\pi_{\bar{\psi}^\alphad}^I&=-(\pi_{\psi^\alpha}^I)^\dagger \ .
\end{align}
As in Sec.~\ref{sec:III}, we assume $\det \left(V^{00}_{ab}\right)^{(0)}\neq 0$. 
Substituting \eqref{Mom_phi_multi} into \eqref{Mom_psi_multi} for eliminating $\dot{\phi}^a$, we obtain the $4N$ primary constraints, 
\begin{align}
	\Phi_{\psi^\alpha}^{I}=&\pi_{\psi^\alpha}^I+P^{(2)0,I}_{\alpha}-S^{00,I}_{a\alpha}(V^{00})^{-1\, ab}P^{(1)0}_b
	+S^{00,I}_{a\alpha}(V^{00})^{-1\, ab}\pi_{\phi^b}-\bigl[S^{00,I}_{a\alpha}(V^{00})^{-1\, ab}V^{0i}_{bc}-S^{i0,I}_{c\alpha}\bigr]\partial_i\phi^c\nonumber\\
	&-\bigl[S^{00,I}_{a\alpha}(V^{00})^{-1\, ab}S^{0i,J}_{b\beta}+W^{0i,IJ}_{\alpha\beta}\bigr]\partial_i\psi^\beta_J+\partial_i\bar{\psi}^{\betad}_J\bigl[S^{00,I}_{a\alpha}(V^{00})^{-1\, ab}(S^{0i,J}_{b\beta})^\dagger+Q^{i0,JI}_{\alpha\betad}\bigr]\ ,\nonumber\\
	\Phi_{\bar{\psi}^\alphad}^I=&-\bigl(\Phi_{\psi^\alpha}^I\bigr)^\dagger \ .
\end{align}

%%%%%%%%%%%%%%%%%%%%%%%%%%%%%%%%%%%%%
\section{A systematic way to find individual functions}
\label{app:D}
%%%%%%%%%%%%%%%%%%%%%%%%%%%%%%%%%%%%%
We show a systematic way to construct the concrete expression for the coefficients of the derivatives in \eqref{Lag_multi}, while in the case of $N=1$, we have constructed them one by one. 
Our prescription proposes a minimal set of functions categorized with respect to each tensorial type, but we do not, at this moment, consider the elimination of the redundant functions of which the origin is in the invariance of a Lagrangian under the addition of total derivatives when the proposed functions are used as the coefficients in the Lagrangian. In Sec.~\ref{subsec:linear}, we concretely reduced the redundant functions coming from the total derivatives in the $N=1$ case. We also leave the introduction of the symmetries~\eqref{Prop_coef} here.
Let us first introduce a new notation. 
Since any coefficients, in general, have space-time indices and Weyl indices, we label each with
\begin{align}
	({\rm indices} \: {\rm for} \: {\rm space\mathchar`-time}\,|\,{\rm indices }\: {\rm for}\: {\rm Weyl} \:{\rm components}\,) \ ,
\end{align}
according to the tensorial type. We do not include the labels of scalar fields, $a, b,\cdots$,  and fermions, $I, J\cdots$, in the notation since they are not related to any symmetry so far and are supposed to be arbitrarily contracted with functions of $\phi^a$, $\Psi_{IJ}$, and $\bar{\Psi}_{IJ}$. With this notation, we can classify the candidates for the coefficients with respect to the tensorial type, e.g., 
$\{\eta^{\mu\nu}\psi_{\alpha,I}, (\sigma^{\mu\nu}\varepsilon)_{\alpha\beta}\psi^\beta_{I}\}\subset(\mu\nu|\alpha)$. 
Then, $S^{\mu\nu, I}_{a\alpha}$ should be composed of $(\mu\nu|\alpha)$ tensors, and similar statements also apply to other coefficients as 
\begin{align}
	&P_0\in (|) \ , \nonumber\\
	&P^{(1)}{}_a^\mu \in (\mu|) \ , \quad  P^{(2)}{}_\alpha^{\mu, I} \in (\mu|\alpha)\ ,\nonumber\\
	&V^{\mu\nu}_{ab}\in (\mu\nu|)\ , \quad S^{\mu\nu, I}_{a\alpha}\in (\mu\nu|\alpha) \ , \quad  W^{\mu\nu,IJ}_{\alpha\beta} \in (\mu\nu|\alpha\beta)\ , \quad Q^{\mu\nu,IJ}_{\alpha\betad} \in (\mu\nu|\alpha\betad)\ .
	\label{Coeff_type}
\end{align}
Thus, our goal in this Appendix is to find the most general form of \eqref{Coeff_type}. 
Let us begin with focusing on the quantities without the Weyl fields as well as the scalar fields, given by
\begin{align}
	&\text{no space-time index}: \{1,  \varepsilon_{\alpha\beta}, \varepsilon_{\alphad\betad}\} \ , \label{tensor_type_00}\\
	&\text{one space-time index}: \{\sigma_{\alpha\alphad}^\mu\}\ \label{tensor_type_10},\\
	&\text{two space-time indices}: \{ \eta^{\mu\nu}, (\sigma^{\mu\nu}\varepsilon)_{\alpha\beta}, (\varepsilon\bar{\sigma}^{\mu\nu})_{\alphad\betad}, 
	 (\sigma^{\rho\mu}\varepsilon)_{\alpha\beta}(\varepsilon\bar{\sigma}^{\rho\nu})_{\gammad\deltad}\} \ .\label{tensor_type_20}
\end{align}
Here, we considered tensors with up to two space-time indices, which are needed to construct \eqref{Coeff_type}, and classified them according to the number of the space-time indices. 
Other tensors can be obtained by combining and contracting elements of \eqref{tensor_type_10} and \eqref{tensor_type_20}, e.g., $\sigma_{\nu\bullet\bullet} (\sigma^{\mu\nu}\varepsilon)_{\bullet\bullet}$. It is, however, shown by using \eqref{B3}--\eqref{A11} that any combination of tensors in \eqref{tensor_type_10} and \eqref{tensor_type_20} with any contraction of space-time indices is reduced to the linear combination of \eqref{tensor_type_10} and \eqref{tensor_type_20} with coefficients of \eqref{tensor_type_00}, just as $\sigma_{\nu\bullet\bullet} (\sigma^{\mu\nu}\varepsilon)_{\bullet\bullet}$ is a linear combination of $\eps_{\bullet\bullet}\sigma^\mu_{\bullet\bullet}$ according to \eqref{A7}. 
What we should consider next is the contractions of Weyl indices with $\varepsilon^{\alpha\beta}$ and $\varepsilon^{\alphad\betad}$ for each combination of elements in \eqref{tensor_type_00}--\eqref{tensor_type_20}. If the contraction is taken for an index of \eqref{tensor_type_00}, it does not produce new types of tensors; e.g., $\varepsilon_{\alpha\beta}\sigma^\mu_{\alpha_1\bullet}\times\varepsilon^{\alpha\alpha_1}$ is just reduced to $-\delta^{\alpha_1}_\beta \sigma^\mu_{\alpha_1\bullet}$. The contraction of two Weyl indices of each element in \eqref{tensor_type_20} vanishes as they are symmetric.
Therefore, in order to find the most general form of \eqref{Coeff_type}, 
Weyl indices of each combination of elements in \eqref{tensor_type_00}--\eqref{tensor_type_20} should not be contracted with epsilon tensors but with the Weyl fields~$\psi^\alpha$ and $\psib^\alphad$.

We begin with the quantities with no space-time index such as 
$(|\alpha), (|\alphad), (|\alpha\beta), (|\alpha\betad)$, and $(|\alphad\betad)$.
These are simply obtained by multiplying the building block \eqref{tensor_type_00} by $\psi^\alpha$ and $\psib^\alphad$.
All the possible combinations for each are listed in Table~\ref{Table1}. 
[Column 1 is composed of an element in \eqref{tensor_type_00} and the Weyl fields, and a cell in column 2 is made up of the combination of the upper left elements.]
The quantities only with one space-time index such as
$(\mu|)$, $(\mu|\alpha)$, and $(\mu|\alphad)$ can be similarly obtained by multiplying the building block \eqref{tensor_type_10} by the Weyl fields, shown in Table~\ref{Table2}. [A cell in column 2 is composed of the combination of the upper left element and $(|\alpha)$, $(|\alphad)$.]
In a similar way, Table~\ref{Table3} is obtained from the building block \eqref{tensor_type_20} as the list of the quantities with two space-time indices.
[A cell in column 3 is composed of the combination of the upper left elements with no Weyl indices and $(|\alpha\beta)$, $(|\alpha\betad)$, $(|\alphad\betad)$.]

We can write down all the coefficients in \eqref{Lag_multi} with the linear combinations of the elements listed in Tables~\ref{Table1}--\ref{Table3} according to each tensorial type.

\begin{table}[h]
\begin{minipage}[t]{.3\textwidth}
\caption{For the sequence with no space-time index.}
\begin{align}
\begin{array}[t]{l|c|c}
{\rm Type}& {\rm Column}\: {\rm 1}& {\rm Column}\: {\rm 2} \\ \hline\hline
(|\alpha) & \psi_{\alpha, I} &\\ 
(|\alphad) & \psib_{\alphad,I} &  \\ \hline
(|\alpha\beta) & \eps_{\alpha\beta} & \psi_{\alpha, I} \psi_{\beta, J}\\
(|\alpha\betad) &&  \psi_{\alpha, I} \psib_{\betad, J}\\
(|\alphad\betad) & \eps_{\alphad\betad} & \psib_{\alphad, I} \psib_{\betad, J}
\end{array} \nonumber
\end{align}
\label{Table1}
\end{minipage}
\begin{minipage}[t]{.6\textwidth}
\caption{For the sequence with one space-time index.}
\begin{equation}
\begin{array}[t]{l|l|l}
{\rm Type}& {\rm Column}\: {\rm 1}& {\rm Column}\: {\rm 2} \\ \hline\hline
(\mu|)& J^\mu_{IJ}\\ \hline
(\mu|\alpha)&\sigma_{\alpha\betad}^\mu\bar{\psi}^\betad_I & J^\mu_{IJ}\psi_{\alpha K}\\
(\mu|\alphad)&\sigma_{\beta\alphad}^\mu\psi^\beta_I &  J^{\mu}_{IJ}\bar{\psi}_{\alphad  K}
\end{array}\nonumber
\end{equation}
\label{Table2}
\end{minipage}
\end{table}
\begin{table}[h]
	\caption{For the sequence with two space-time indices.}
	\begin{align}
		&\begin{array}[t]{l|l|l|l}
			{\rm Type}& {\rm Column}\: {\rm 1}& {\rm Column}\: {\rm 2} & \\ \hline\hline
			(\mu\nu|)& \eta^{\mu\nu} , \, K^{\mu\nu}_{IJ} , \, \bar{K}^{\mu\nu}_{IJ} , \, K^{\mu\rho}_{IJ}\bar{K}_\rho{}^\nu{}_{KL} & \\ \hline
			(\mu\nu|\alpha)& (\sigma^{\mu\nu}\varepsilon)_{\alpha\beta}\psi^\beta_I , \, (\sigma^{\mu\rho}\varepsilon)_{\alpha\beta}\psi^{\beta}_I\bar{K}_\rho{}^\nu{}_{JK} & \eta^{\mu\nu}\psi_{\alpha I} , \, K^{\mu\nu}_{IJ}\psi_{\alpha K} , \, \bar{K}^{\mu\nu}_{IJ}\psi_{\alpha K} , \, K^{\mu\rho}_{IJ}\bar{K}_\rho{}^\nu{}_{KL}\psi_{\alpha M}\\
			(\mu\nu|\alphad)& (\varepsilon\bar{\sigma}^{\mu\nu})_{\alphad\betad}\bar{\psi}^\betad_{I} , \, K^{\mu\rho}_{IJ}(\varepsilon\bar{\sigma}_\rho{}^\nu)_{\alphad\betad}\bar{\psi}^\betad_K &  \eta^{\mu\nu}\bar{\psi}_{\alphad  I} , \, K^{\mu\nu}_{IJ}\bar{\psi}_{\alphad  K} , \, \bar{K}^{\mu\nu}_{IJ}\bar{\psi}_{\alphad  K} , \, K^{\mu\rho}_{IJ}\bar{K}_\rho{}^\nu{}_{KL}\bar{\psi}_{\alphad M}\\ \hline
			(\mu\nu|\alpha\beta)& (\sigma^{\mu\nu}\varepsilon)_{\alpha\beta} , \, (\sigma^{\mu\rho}\varepsilon)_{\alpha\beta}\bar{K}_\rho{}^\nu{}_{IJ} & (\sigma^{\mu\nu}\varepsilon)_{\alpha\gamma}\psi^\gamma_I\psi_{\beta J} , \, (\sigma^{\mu\rho}\varepsilon)_{\alpha\gamma}\psi^{\gamma}_I\bar{K}_\rho{}^\nu{}_{JK}\psi_{\beta L}, \, {(\sigma^{\mu\nu}\varepsilon)_{\beta\gamma}\psi^\gamma_I\psi_{\alpha J} , \, (\sigma^{\mu\rho}\varepsilon)_{\beta\gamma}\psi^{\gamma}_I\bar{K}_\rho{}^\nu{}_{JK}\psi_{\alpha L}} &\\ 
			(\mu\nu|\alpha\betad) & (\sigma^{\mu\rho}\varepsilon)_{\alpha\gamma}\psi^{\gamma}_I(\varepsilon\bar{\sigma}_\rho{}^\nu)_{\betad\gammad}\bar{\psi}^\gammad_J & (\sigma^{\mu\nu}\varepsilon)_{\alpha\gamma}\psi^\gamma_I\bar{\psi}_{\betad J} , \, (\sigma^{\mu\rho}\varepsilon)_{\alpha\gamma}\psi^{\gamma}_I\bar{K}_\rho{}^\nu{}_{JK}\bar{\psi}_{\betad L} ,\, (\varepsilon\bar{\sigma}^{\mu\nu})_{\betad\gammad}\bar{\psi}^\gammad_{I}\psi_{\alpha J} , \, K^{\mu\rho}_{IJ}(\varepsilon\bar{\sigma}_\rho{}^\nu)_{\betad\gammad}\bar{\psi}^\gammad_K \psi_{\alpha L} \\
			(\mu\nu|\alphad\betad) & (\varepsilon\bar{\sigma}^{\mu\nu})_{\alphad\betad} , \, K^{\mu\rho}_{IJ}(\varepsilon\bar{\sigma}_\rho{}^\nu)_{\alphad\betad} & (\varepsilon\bar{\sigma}^{\mu\nu})_{\alphad\gammad}\psib^\gammad_I\psib_{\betad J}, \, K^{\mu\rho}_{IJ}(\varepsilon\bar{\sigma}{}_\rho{}^\nu)_{\alphad\gammad}\psib^\gammad_K\psib_{\betad L}, \, {(\varepsilon\bar{\sigma}^{\mu\nu})_{\betad\gammad}\psib^\gammad_I\psib_{\alphad J}, \, K^{\mu\rho}_{IJ}(\varepsilon\bar{\sigma}{}_\rho{}^\nu)_{\betad\gammad}\psib^\gammad_K\psib_{\alphad L}}&\\
		\end{array} \nonumber\\ 
		&\qquad \qquad \begin{array}[t]{l|l|l|l}
			& {\rm Column}\: {\rm 3} \\ \hline\hline
			& \\ \hline
			& \\ 
			& \\ \hline
			& \eta^{\mu\nu}\psi_{\alpha I}\psi_{\beta J} , \, K^{\mu\nu}_{IJ}\psi_{\alpha K}\psi_{\beta L} , \, \bar{K}^{\mu\nu}_{IJ}\psi_{\alpha K}\psi_{\beta L} , \, K^{\mu\rho}_{IJ}\bar{K}_\rho{}^\nu{}_{KL}\psi_{\alpha M}\psi_{\beta N} , \, \eta^{\mu\nu}\varepsilon_{\alpha\beta} , \, K^{\mu\nu}_{IJ}\varepsilon_{\alpha\beta} , \, \bar{K}^{\mu\nu}_{IJ}\varepsilon_{\alpha\beta} , \, K^{\mu\rho}_{IJ}\bar{K}_\rho{}^\nu{}_{KL}\varepsilon_{\alpha\beta}\\ 
			& \eta^{\mu\nu}\psi_{\alpha I}\bar{\psi}_{\betad J} , \, K^{\mu\nu}_{IJ}\psi_{\alpha K}\bar{\psi}_{\betad L} , \, \bar{K}^{\mu\nu}_{IJ}\psi_{\alpha K}\bar{\psi}_{\betad L} , \, K^{\mu\rho}_{IJ}\bar{K}_\rho{}^\nu{}_{KL}\psi_{\alpha M}\bar{\psi}_{\betad N}\\
			& \eta^{\mu\nu}\psib_{\alphad I}\psib_{\betad J} , \, K^{\mu\nu}_{IJ}\psib_{\alphad K}\psib_{\betad L} , \, \bar{K}^{\mu\nu}_{IJ}\psib_{\alphad K}\psib_{\betad L} , \, K^{\mu\rho}_{IJ}\bar{K}_\rho{}^\nu{}_{KL}\psib_{\alphad M}\psib_{\betad N} , \, \eta^{\mu\nu}\varepsilon_{\alphad\betad} , \, K^{\mu\nu}_{IJ}\varepsilon_{\alphad\betad} , \, \bar{K}^{\mu\nu}_{IJ}\varepsilon_{\alphad\betad} , \, K^{\mu\rho}_{IJ}\bar{K}_\rho{}^\nu{}_{KL}\varepsilon_{\alphad\betad}\\
		\end{array} \nonumber
	\end{align}
	\label{Table3}
\end{table}

%---------   References   ---------%
%

%\bibliography{reference_gen_d}

\end{document}